\documentstyle[11pt,aip,equations,epsf,times,hyperref]{article}
\def\bibliostyle{report}

\ifx\undefined\SetMathAlphabet
\let\mathsf=\sf
\let\mathbf=\bf
\else
\SetMathAlphabet{\mathsf}{normal}{\encodingdefault}{\sfdefault}%
	{\bfdefault}{n}
\fi
\def~{\penalty200 \ }

\def\abs#1{\left|#1\right|}
\def\sfrac#1#2{{\textstyle\frac{#1}{#2}}}

\def\s{{*}}
\def\suba#1#2{_{[#1]#2}}
\def\subb#1#2#3{_{#1[#2]#3}}
\def\ssuba#1#2#3{_{{#1}[#2]#3}}
\def\ssubb#1#2#3#4{_{#1#2[#3]#4}}
\def\mat[#1,#2,#3]%
    {\left[\!\!
      \begin{array}{c}\struta#1\\\struta#2\\\struta#3\end{array}
    \!\!\right]}
\def\II{{\mathsf I}}
\def\UU{{\mathsf U}}
\def\DD{{\mathsf D}}
\def\LL{{\mathsf L}}
\def\F{{\mathbf F}}
\def\K{{\mathbf K}}
\def\u{{\mathbf u}}

\def\d{\partial}

\def\pd#1#2{\mathchoice{\d#1\over\d#2}{\d#1/\d#2}{\d#1/\d#2}{\d#1/\d#2}}
\def\pdd#1#2{\mathchoice{\d^2#1\over\d#2^2}%
  {\d^2#1/\d#2^2}{\d^2#1/\d#2^2}{\d^2#1/\d#2^2}}
\def\maxw{m}		%{,{\rm max}}
\def\jb#1#2{\tilde\jmath_{#1,#2}}

\def\qb#1#2{\tilde q_{#1,#2}}
\def\sigman{\bar\sigma}

\begin{document}
\date{PPPL--2598 (1989) [Phys.~Fluids B {\bf 1}, 1355--1368 (1989)]}
\author{Bastiaan J. Braams and Charles F. F. Karney\\
Plasma Physics Laboratory\\
Princeton University\\
Princeton, NJ 08543-0451}
\title{Conductivity of a Relativistic Plasma}
\maketitle
\begin{abstract}
The collision operator for a relativistic plasma is reformulated in terms
of an expansion in spherical harmonics.  In this formulation the collision
operator is expressed in terms of five scalar potentials which are given by
one-dimensional integrals over the distribution function.  This formulation
is used to calculate the electrical conductivity of a uniform electron-ion
plasma with infinitely massive ions.
\end{abstract}

\section{Introduction}

Landau \cite{Landau36} first obtained an accurate kinetic equation for a
nonrelativistic plas\-ma. The Landau collision operator was generalized to
a relativistic plasma by Beliaev and Budker \cite{BelBud56}.  The collision
operator in both cases involves integrals of the distribution function of
the background species.  This makes the operators difficult to evaluate,
numerically or analytically.

In the nonrelativistic case this difficulty was removed by Rosenbluth,
MacDonald, and Judd \cite{RMJ57}, and by Trubnikov \cite{Trubni58}.  They
recast the Landau operator into differential form by writing it in terms of
derivatives of two scalar potentials. The potentials in turn satisfy a pair of
elliptic partial differential equations.  With the aid of this formulation,
the numerical evaluation of the collision operator may be accomplished
straightforwardly by solving the potential equations.  An analytical solution
of these equations in terms of spherical harmonics was given by Rosenbluth
{\it et al}\ \cite{RMJ57}.

Recently \cite{BraKar87}, we formulated the relativistic collision operator
of Beliaev and Budker \cite{BelBud56} in terms of five scalar potentials,
which again obey elliptic partial differential equations.  In the present
paper we extend this formulation by solving the potential equations in
terms of an expansion in spherical harmonics.  As an application, we
calculate the electrical conductivity of a relativistic plasma with massive
ions.

In Sec.~\ref{potentials} we review the differential formulation of the
collision operator.  The spherical harmonic expansion is developed in
Sec.~\ref{solution}.  It is shown in Sec.~\ref{nonrelativistic} how
the nonrelativistic results are recovered in the limit $c\to\infty$.
We evaluate the zeroth-order and first-order spherical harmonic components
of the collision term with respect to a Maxwellian background in
Secs.~\ref{isotropic} and \ref{first-harmonic}.  The calculation of the
conductivity is given in Sec.~\ref{conductivity}.

\section{Potentials}\label{potentials}

We begin by summarizing the differential formulation of the collision operator.
This repeats our earlier exposition \cite{BraKar87}; however, we now introduce
a more compact notation and also write the results in such a way that the
nonrelativistic limit is more easily obtained.

The collision term for a plasma of species $s$ colliding off species $s'$
may be written in the Fokker-Planck form as \cite{Landau36,BelBud56}:
\begin{equation}\label{coll-op}
C^{s/s'}(f_s,f_{s'})=
\pd{}{\u}\cdot\biggl(\DD^{s/s'}\cdot\pd{f_s}{\u}-\F^{s/s'}f_s\biggr),
\end{equation}
where the diffusion and friction coefficients are given by
\begin{subequations}\label{df-def}
\begin{eqalignno}
\DD^{s/s'}(\u)&=\frac{\Gamma^{s/s'}}{2n_{s'}}
  \int\UU(\u,\u')f_{s'}(\u')\,d^3\u',\\
\F^{s/s'}(\u)&=-\frac{\Gamma^{s/s'}}{2n_{s'}}\frac{m_s}{m_{s'}}
  \int\biggl(\pd{}{\u'}\cdot\UU(\u,\u')\biggr)f_{s'}(\u')\,d^3\u',\\
\Gamma^{s/s'}&=\frac{n_{s'}q_s^2q_{s'}^2\log\Lambda^{s/s'}}
  {4\pi\epsilon_0^2m_s^2}.\nonumber
\end{eqalignno}
\end{subequations}
Here, $\u$ is the momentum per unit rest mass, and, in the relativistic
case, the kernel $\UU$ is given by \cite{BelBud56}
\begin{equation}\label{kernel}
\UU(\u,\u')=\frac{r^2}{\gamma\gamma'w^3}
  \bigl(w^2\II-\u\u-\u'\u'+r(\u\u'+\u'\u)\bigr),
\end{equation}
in which $\gamma=\sqrt{1+\u^2/c^2}$,
$\gamma'=\sqrt{1+\u'^2/c^2}$, and
\begin{subequations}\label{rw-def}
\begin{eqalignno}
  r&=\gamma\gamma'-\u\cdot\u'/c^2,\\
  w&=c\sqrt{r^2-1}.
\end{eqalignno}
\end{subequations}
The quantity $r$ is the relativistic correction factor corresponding to the
relative velocity of the two interacting particles.  The relative speed of the
interacting particles is given by $w/r$.  In the nonrelativistic limit,
$r\to1$ and $w\to\abs{\u-\u'}$.

In the previous paper \cite{BraKar87} we expressed the Eqs.~(\ref{df-def})
in differential form, making use of the potentials
\begin{subequations}\label{pot-int-prl}
\def\dw{\frac{d^3\u'}{\gamma'}}
\begin{eqalignno}
  \Psi\ssuba{s'}10(\u)&=-\frac1{4\pi}\int w^{-1}f_{s'}(\u')\dw,\\
  \Psi\ssuba{s'}2{02}(\u)&=-\frac1{8\pi}\int wf_{s'}(\u')\,\dw,\\
  \Psi\ssuba{s'}3{022}(\u)&=-\frac1{32\pi}
    \int c^3\bigl(r\sinh^{-1}(w/c)-w/c\bigr)f_{s'}(\u')\,\dw,\\
  \Psi\ssuba{s'}11(\u)&=-\frac1{4\pi}\int rw^{-1}f_{s'}(\u')\,\dw,\\
  \Psi\ssuba{s'}2{11}(\u)&=-\frac1{8\pi}
    \int c\sinh^{-1}(w/c)f_{s'}(\u')\,\dw.
\end{eqalignno}
\end{subequations}
These potentials satisfy the differential equations
\begin{subequations}\label{pot-diff-prl}
\begin{eqalignno}
L_0\Psi\ssuba{s'}10&=f_{s'},\\
L_2\Psi\ssuba{s'}2{02}&=\Psi\ssuba{s'}10,\\
L_2\Psi\ssuba{s'}3{022}&=\Psi\ssuba{s'}2{02},\\
L_1\Psi\ssuba{s'}11&=f_{s'},\\
L_1\Psi\ssuba{s'}2{11}&=\Psi\ssuba{s'}11,
\end{eqalignno}
\end{subequations}
where
\begin{equation}\label{L-def}
L_a\Psi=\biggl(\II+\frac{\u\u}{c^2}\biggr):\frac{\d^2\Psi}{\d\u\d\u}
  +\frac{3\u}{c^2}\cdot\pd\Psi\u + \frac{1-a^2}{c^2}\Psi.
\end{equation}
In terms of these potentials the diffusion and friction coefficients are
given by
\begin{subequations}\label{diff-fric}
\begin{eqalignno}
\DD^{s/s'}(\u)&=-\frac{4\pi\Gamma^{s/s'}}{n_{s'}}
  \biggl[\frac1{\gamma}\biggl(\LL+\frac{\II}{c^2}+\frac{\u\u}{c^4}\biggr)
  \Psi\ssuba{s'}2{02}\nonumber\\
&\qquad\qquad\qquad\qquad{}-\frac4{\gamma c^2}
\biggl(\LL-\frac{\II}{c^2}-\frac{\u\u}{c^4}\biggr)\Psi\ssuba{s'}3{022}\biggr],
  \label{diffusion-eq}\\
\F^{s/s'}(\u)&=-\frac{4\pi\Gamma^{s/s'}}{n_{s'}}\frac{m_s}{m_{s'}}
  \frac1{\gamma}\biggl[\K \Psi\ssuba{s'}11-
  \frac2{c^2}\K \Psi\ssuba{s'}2{11}\biggr],\label{friction-eq}
\end{eqalignno}
\end{subequations}
where
\begin{eqalignno*}
\LL\Psi(\u)&=\biggl(\II+\frac{\u\u}{c^2}\biggr)\cdot\frac{\d^2\Psi}{\d\u\d\u}
  \cdot\biggl(\II+\frac{\u\u}{c^2}\biggr)
  +\biggl(\II+\frac{\u\u}{c^2}\biggr)\biggl(\u\cdot\pd\Psi\u\biggr),\\
\K\Psi(\u)&=\biggl(\II+\frac{\u\u}{c^2}\biggr)\cdot\pd\Psi\u.
\end{eqalignno*}
Equations~(\ref{pot-diff-prl}) and~(\ref{diff-fric}) constitute the
differential formulation of the relativistic collision operator. Boundary
conditions on the solutions of Eqs.~(\ref{pot-diff-prl}) are obtained by
expansion of Eqs.~(\ref{pot-int-prl}) for $|\u|\to\infty$.

The present notation differs from that in Ref.~\citenum{BraKar87}.
The correspondence is
$\Psi\ssuba{s'}10 \linebreak[0] \equiv h_0$,
$\Psi\ssuba{s'}2{02} \equiv h_1$,
$\Psi\ssuba{s'}3{022} \equiv h_2$,
$\Psi\ssuba{s'}11 \equiv g_0$,
$\Psi\ssuba{s'}2{11} \equiv g_1$.
The present notation reflects more clearly the structure of the differential
equations~(\ref{pot-diff-prl}). By adjoining to Eqs.~(\ref{pot-int-prl})
the definition
$$
\Psi\ssuba{s'}0{}(\u)=f_{s'}(\u),
$$
and by using ``$\s$'' in the context $\Psi\ssuba{s'}k\s$ to stand for a string
of $k\ge 0$ indices, we can express the five potential equations
(\ref{pot-diff-prl}) in the concise form
\begin{equation}\label{pot-diff}
L_a\Psi\ssuba{s'}{k+1}{\s a}=\Psi\ssuba{s'}k\s.
\end{equation}
The integral representations~(\ref{pot-int-prl}) will likewise condense
into a single formula
\begin{equation}\label{pot-int}
\Psi\ssuba{s'}k\s(\u)=\frac1{4\pi}\int c^{2k-3}y\subb 0k\s(w/c)f_{s'}(\u')
  \frac{d^3\u'}{\gamma'},
\end{equation}
for $k>0$.  The kernel function $y\subb 0k\s$ is defined in the next section.

\section{Solution to potential equations}\label{solution}

In this section, Eq.~(\ref{pot-diff}) will be solved by separation of
variables in spherical coordinates.  The choice of spherical coordinates is
a natural one since frequently the distribution function is nearly
spherically symmetric and so is well represented by only a few spherical
harmonics.  Since $\Psi\ssuba sk\s$ depends on the distribution of species
$s$ only, we will simplify the notation by dropping the species subscripts.

\subsection{Spherical harmonic expansion}

In a spherical $(u,\theta,\phi)$ coordinate system, the operator $L_a$ is
\begin{eqalignno}\label{L-comp}
L_a\Psi&=\gamma^2\pdd\Psi{u}+\biggl(\frac2u+\frac{3u}{c^2}\biggr)\pd\Psi{u}+
  \frac1{u^2}\biggl(\pdd\Psi\theta+\cot\theta\pd\Psi\theta\biggr)\nonumber\\
&\qquad\qquad\qquad+\frac1{u^2\sin^2\theta}\pdd\Psi\phi+\frac{1-a^2}{c^2}\Psi.
\end{eqalignno}
Let us expand the potentials in terms of Legendre harmonics:
\begin{equation}\label{spher-harm}
\Psi\suba k\s(\u)=\sum_{l=0}^\infty\sum_{m=-l}^l
  \psi\subb {lm}k\s(u)P_l^m(\cos\theta)\exp(im\phi).
\end{equation}
The coefficients $\psi\subb {lm}k\s(u)$ are given by
$$
\psi\subb {lm}k\s(u) = (2l+1)\frac{(l-m)!}{(l+m)!}
\int \frac{d\Omega}{4\pi} \Psi\suba k\s(\u)
P_l^m(\cos\theta)\exp(-im\phi),
$$
where $d\Omega$ is an element of solid angle
$$
\int d\Omega\,\ldots =
\int_0^\pi\!\! \sin\theta\, d\theta \int_0^{2\pi}\!\! d\phi\,\ldots.
$$
Equation~(\ref{pot-diff}) becomes
\begin{equation}\label{spher-eq}
L_{l,a}\psi\subb{lm}{k+1}{\s a}=\psi\subb{lm}k\s,
\end{equation}
where
\begin{equation}\label{spher-op}
L_{l,a}\chi(u)=\biggl(1+\frac{u^2}{c^2}\biggr)\frac{d^2\chi}{du^2}
+\biggr(\frac 2u+\frac{3u}{c^2}\biggl)\frac{d\chi}{du}
  -\biggl(\frac{l(l+1)}{u^2}+\frac{a^2-1}{c^2}\biggr)\chi,
\end{equation}
and
where $\psi\subb {lm}0{}(u) = f_{lm}(u)$ is the $(l,m)$ coefficient in the
expansion of $f(\u)$ in Legendre harmonics.

As before, Eq.~(\ref{spher-eq}) stands for a set of five differential
equations for the potentials $\psi\subb{lm}10$, $\psi\subb{lm}2{02}$,
$\psi\subb{lm}3{022}$, $\psi\subb{lm}11$, and $\psi\subb{lm}2{11}$.  We
will, however, find it convenient to solve for the potentials with
arbitrary indices, i.e., to solve the system of equations
\begin{subequations}\label{inhomo}
\begin{eqalignno}
L_{l,a}\psi\subb{lm}1a&=f_{lm},\\
L_{l,a'}\psi\subb{lm}2{aa'}&=\psi\subb{lm}1a,\\
L_{l,a''}\psi\subb{lm}3{aa'a''}&=\psi\subb{lm}2{aa'},
\end{eqalignno}
\end{subequations}
for arbitrary $a$, $a'$, and $a''$.

For later reference we list the components of the operators $\LL$ and $\K$
that occur in Eqs.~(\ref{diff-fric}):
\begin{subequations}\label{lk-comp}
\begin{eqalignno}
  \LL_{uu}\Psi&=\gamma^4\pdd\Psi{u}+\frac{\gamma^2u}{c^2}\pd\Psi{u},\\
  \LL_{\theta\theta}\Psi&=\frac1{u^2}\pdd\Psi\theta+
    \frac{\gamma^2}u\pd\Psi{u},\\
  \LL_{\phi\phi}\Psi&=\frac1{u^2\sin^2\theta}\pdd\Psi\phi+
    \frac{\gamma^2}u\pd\Psi{u}+\frac1{u^2}\cot\theta\pd\Psi\theta,\\
  \LL_{u\theta}\Psi&=\LL_{\theta u}\Psi=
    \frac{\gamma^2}{u}\biggl(\pd{^2\Psi}{u\d\theta}-
      \frac1{u}\pd\Psi\theta\biggr),\\
  \LL_{u\phi}\Psi&=\LL_{\phi u}\Psi=
    \frac{\gamma^2}{u\sin\theta}\biggl(\pd{^2\Psi}{u\d\phi}-
      \frac1{u}\pd\Psi\phi\biggr),\\
  \LL_{\theta\phi}\Psi&=\LL_{\phi\theta}\Psi=
    \frac1{u^2\sin\theta}\biggl(\pd{^2\Psi}{\theta\d\phi}-
      \cot\theta\pd\Psi\phi\biggr),
\end{eqalignno}
\begin{eqalignno}
  \K_{u}\Psi&=\gamma^2\pd\Psi{u},\\
  \K_{\theta}\Psi&=\frac1u\pd\Psi\theta,\\
  \K_{\phi}\Psi&=\frac1{u\sin\theta}\pd\Psi\phi,
\end{eqalignno}
\begin{equation}
\biggl(\II+\frac{\u\u}{c^2}\biggr)\Psi=\left(
\begin{array}{ccc}
\gamma^2 & 0 & 0 \\ 0 & 1 & 0 \\ 0 & 0 & 1
\end{array}
\right)\Psi.
\end{equation}
\end{subequations}

\subsection{Homogeneous solutions}\label{homogeneous}

In order to solve the inhomogeneous equations~(\ref{inhomo}), it is
required first to determine the solutions to the homogeneous equations
\begin{subequations}\label{homo}
\begin{eqalignno}
L_{l,a}\psi^{\rm HS}\subb{lm}1a&=0,\label{homo-a}\\
L_{l,a'}\psi^{\rm HS}\subb{lm}2{aa'}&=\psi^{\rm HS}\subb{lm}1a,\\
L_{l,a''}\psi^{\rm HS}\subb{lm}3{aa'a''}&=\psi^{\rm HS}\subb{lm}2{aa'}.
\end{eqalignno}
\end{subequations}

It is shown in the appendix that two independent solutions to
Eq.~(\ref{homo-a}) are $\psi^{\rm HS}\subb{lm}1a(u)\linebreak[0]=j\subb l1a(u/c)$ and
$\psi^{\rm HS}\subb{lm}1a(u)=y\subb l1a(u/c)$ where
\begin{subequations}\label{jy-def}
\begin{eqalignno}
j\subb l1a(u/c)&=\sqrt{\frac{\pi c}{2u}}P^{-l-1/2}_{a-1/2}(\gamma),
  \label{j-def}\\
y\subb l1a(u/c)&=(-1)^{-l-1}\sqrt{\frac{\pi c}{2u}}P^{l+1/2}_{a-1/2}(\gamma),
  \label{y-def}
\end{eqalignno}
\end{subequations}
and $P^\mu_\nu$ is the associated Legendre function of the first kind.

When the full system of equations (\ref{homo}) is considered, we need to
introduce the additional functions, defined recursively by
Eq.~(\ref{high-def}),
\begin{equation}\label{high-def-rep}
j\subb l{k+2}{\s aa'} =
\begin{cases}
  \displaystyle\frac{j\subb l{k+1}{\s a}-j\subb l{k+1}{\s a'}}
    {a^2-a'^2},&for $a\ne a'$,\\
  \displaystyle\pd{j\subb l{k+1}{\s a}}{(a^2)},
    &for $a=a'$,
\end{cases}
\end{equation}
with $y\subb lk\s$ defined in a similar fashion.  The general solution to
Eqs.~(\ref{homo}) is given by
\begin{eqalignno}\label{homo-sol}
\mat[\psi^{\rm HS}\subb{lm}1a,\psi^{\rm HS}\subb{lm}2{aa'},
\psi^{\rm HS}\subb{lm}3{aa'a''}]&=
C\subb{lm}1a \mat[j\subb l1a,c^2 j\subb l2{aa'},c^4 j\subb l3{aa'a''}]
+C'\subb{lm}1a \mat[y\subb l1a,c^2 y\subb l2{aa'},c^4 y\subb l3{aa'a''}]
\nonumber\\*&\qquad
+C\subb{lm}2{aa'} \mat[0,j\subb l1{a'},c^2 j\subb l2{a'a''}]
+C'\subb{lm}2{aa'} \mat[0,y\subb l1{a'},c^2 y\subb l2{a'a''}]
\nonumber\\*&\qquad
+C\subb{lm}3{aa'a''} \mat[0,0,j\subb l1{a''}]
+C'\subb{lm}3{aa'a''} \mat[0,0,y\subb l1{a''}],
\end{eqalignno}
where $C\subb{lm}k\s$ and $C'\subb{lm}k\s$ are arbitrary constants (independent
of $u$) and the argument of the functions $j\subb lk\s$ and $y\subb lk\s$
is $u/c$.

The functions $j\subb lk\s$ and $y\subb lk\s$ are invariant under permutation
of the indices in $\s$ and invariant under change of sign of any index in $\s$.
Also the functions satisfy $j\subb lk\s(-z) = (-1)^lj\subb lk\s(z)$ and
$y\subb lk\s(-z) = (-1)^ly\subb lk\s(z)$; and $j\subb lk\s$ and $y\subb
lk\s$ are related by
$$
y\subb lk\s=(-1)^{l+1}j\subb {-l-1}k\s.
$$
It follows from Eqs.~(\ref{leading-order}) that
$c^{2k+l-2}\* j\subb lk\s(u/c)$ and $c^{2k-l-3}\* y\subb lk\s(u/c)$ reduce to
finite and nonzero expressions in the nonrelativistic limit:
\begin{subequations}\label{jy-nr}
\begin{eqalignno}
\lim_{c\to\infty}c^{2k+l-2}j\subb lk\s(u/c)&=
  \frac{u^{l+2k-2}}{(2k-2)!!(2l+2k-1)!!},\label{j-nr}\\
\lim_{c\to\infty}c^{2k-l-3}y\subb lk\s(u/c)&=
  \frac{(-1)^k(2l-2k+1)!!}{(2k-2)!!u^{l-2k+3}}.\label{y-nr}
\end{eqalignno}
\end{subequations}
Further properties of these functions are given in the appendix.

For our problem, $l$ and the indices $\s$ are always integers, in which
case $j\subb lk\s$ and $y\subb lk\s$ may be expressed in terms of
elementary functions. The functions that we need are given explicitly in
Eqs.~(\ref{l=0}) and~(\ref{l=1}). It is seen there that the kernels
appearing in Eqs.~(\ref{pot-int-prl}) are precisely the functions $y\subb
0k\s$; this justifies the general definition for the potentials given in
Eq.~(\ref{pot-int}).

\subsection{Green's function}

With the solutions of the homogeneous equations in hand, it is a
straightforward matter to construct a Green's function and so to write down
the general solution of the inhomogeneous problem. The final task will be
to apply appropriate boundary conditions from Eq.~(\ref{pot-int}).

We begin by defining the functions
\begin{subequations}\label{N-def}
\begin{eqalignno}
N\subb l0{}(u,u')&=0,\\
N\subb l1a(u,u')&=c^{-1}y\subb l1a(u/c)j\subb l1a(u'/c),\\
N\subb l2{aa'}(u,u')&=c\bigl(y\subb l1a(u/c)j\subb l2{aa'}(u'/c) +
  y\subb l2{aa'}(u/c)j\subb l1{a'}(u'/c)\bigr),\\
N\subb l3{aa'a''}(u,u')&=c^3\bigl(y\subb l1a(u/c)j\subb l3{aa'a''}(u'/c) +
  y\subb l2{aa'}(u/c)j\subb l2{a'a''}(u'/c)\nonumber\\
&\qquad\qquad{}+y\subb l3{aa'a''}(u/c)j\subb l1{a''}(u'/c)\bigr).
\end{eqalignno}
\end{subequations}
From Eqs.~(\ref{jy-nr}), we see that the functions $N\subb lk\s$ reduce to
finite nonzero expressions in the nonrelativistic limit.  These functions
satisfy the differential equations
\begin{eqalignno*}
L_{l,a}N\subb l{k+1}{\s a}(u,u') &= N\subb l{k}\s(u,u'),\\
L_{l,a}N\subb l{k+1}{\s a}(u',u) &= N\subb l{k}\s(u',u).
\end{eqalignno*}

Next we combine $N\subb l{k}\s(u,u')$ and $N\subb l{k}\s(u',u)$ to obtain
the Green's functions
\begin{equation}
  K\subb lk\s(u,u')=
  \begin{cases}
\displaystyle\frac{\gamma}{u^2}\delta(u-u'),&for $k=0$\\
N\subb lk\s(u_>,u_<),&for $k>0$,
  \end{cases}
\end{equation}
where $u_>=\max(u,u')$ and $u_<=\min(u,u')$.
These Green's functions satisfy
$$
L_{l,a}K\subb l{k+1}{\s a}(u,u') = K\subb l{k}\s(u,u').
$$
To establish this relation, we use
$$\left.\pd{}u\bigl(N\subb lk\s(u,u')-N\subb lk\s(u',u)\bigr)\right|_{u=u'}
=
\begin{cases}
\displaystyle \frac1{\gamma u^2}, & for $k=1$,\\
0, & otherwise,
\end{cases}
$$
where the result for $k=1$ follows from the expression for the Wronskian,
Eq.~(\ref{wronskian}).
Furthermore, the Green's functions satisfy equations analogous to
Eqs.~(\ref{high-def-rep}), namely,
\begin{equation}\label{high-K-def}
\frac{K\subb l{k+2}{\s aa'}}{c^2} =
\begin{cases}
  \displaystyle\frac{K\subb l{k+1}{\s a}-K\subb l{k+1}{\s a'}}{a^2-a'^2},
    &for $a\ne a'$,\\
  \displaystyle\pd{K\subb l{k+1}{\s a}}{(a^2)},
    &for $a=a'$.
\end{cases}
\end{equation}
The Green's functions $K\subb lk\s$ are therefore symmetric under interchange
of the indices in $\s$.

A particular integral for the differential equations~(\ref{spher-eq}) is now
given by
\begin{equation}\label{psi-pi}
\psi^{\rm PI}\subb{lm}k\s(u)=
\int_0^\infty K\subb lk\s(u,u') \frac{u'^2}{\gamma'} f_{lm}(u')\,du'.
\end{equation}
We shall show next that this
particular integral is precisely the solution defined by the integral form
of the potential Eq.~(\ref{pot-int}).  This will be done by matching the
behavior of Eq.~(\ref{psi-pi}) near $u=0$ to the behavior of the solutions
of Eq.~(\ref{pot-int}).  For $u\to0$ we have $K\subb lk\s(u,u')\to N\subb
lk\s(u',u)$. By using Eqs.~(\ref{N-def}) and~(\ref{leading-j}) we obtain in
this limit
\begin{equation}\label{pi-lead}
\psi^{\rm PI}\subb{lm}k\s(u)\to
\frac{u^lc^{2k-l-3}}{(2l+1)!!}\int_0^\infty
  y\subb lk\s(u'/c)\frac{u'^2}{\gamma'}f_{lm}(u')\,du'.
\end{equation}

The general solution is obtained by adding the homogeneous solution,
Eq.~(\ref{homo-sol}), to the particular integral: $\psi\subb{lm}k\s =
\psi^{\rm PI}\subb{lm}k\s + \psi^{\rm HS}\subb{lm}k\s$.  It remains to
determine the coefficients $C\subb{lm}k\s$ and $C'\subb{lm}k\s$ appearing
in Eq.~(\ref{homo-sol}).  The primed coefficients $C'\subb{lm}k\s$ are
found by considering just the order of growth of $\psi\subb{lm}k\s(u)$ at
$u=0$. From Eq.~(\ref{spher-harm}) it follows that
$\psi\subb{lm}k\s(u)=O(u^l)$, whereas from Eqs.~(\ref{homo-sol}) and
(\ref{leading-order}), we have $\psi^{\rm HS}\subb{lm}k\s = C'\subb{lm}k\s
O(u^{-l-1})$. The coefficients $C'\subb{lm}k\s$ must vanish in order to
suppress this divergence at $u = 0$.  To establish that the unprimed
coefficients $C\subb{lm}k\s$ also vanish, we must consider more carefully
the behavior of $\psi\subb{lm}k\s(u)$ at the origin.

From Eqs.~(\ref{homo-sol}), (\ref{pi-lead}), and (\ref{leading-j}), we find
that the leading order behavior of $\psi\subb{lm}k\s$ is
\begin{equation}\label{psi-lead}
\psi\subb{lm}k\s \to \biggl[C\subb{lm}k\s +
  c^{2k-3}\int_0^\infty y\subb lk\s(u'/c)\frac{u'^2}{\gamma'}f_{lm}(u')\,du'
\biggr]\frac {(u/c)^l}{(2l+1)!!}.
\end{equation}
In order to determine $C\subb{lm}k\s$, we expand the integral
representation of $\psi\subb{lm}k\s$ near $u=0$.  Substituting the
spherical harmonic expansion, Eq.~(\ref{spher-harm}), into
Eq.~(\ref{pot-int}), we obtain
\begin{equation}\label{pot-int-harm}
\psi\subb{lm}k\s(u)
  = c^{2k-3}\int_0^\infty\!\! u'^2\,du' \int\!\frac{d\Omega}{4\pi}\,
  y\subb 0k\s(w/c) \frac{f_{lm}(u')}{\gamma'}
  \frac{Y_{lm}(\theta',\phi')}{Y_{lm}(\theta,\phi)},
\end{equation}
where we have introduced the spherical harmonics \cite{Jackso75}
$$
Y_{lm}(\theta,\phi)=\sqrt{\frac{2l+1}{4\pi}\frac{(l-m)!}{(l+m)!}}
P_l^m(\cos\theta)\exp(im\phi).$$
Because we want to expand Eq.~(\ref{pot-int-harm}) for small $u$, we
write $w$ from Eq.~(\ref{rw-def}) as
$$w=\sqrt{u'^2\biggl(1+\frac{\epsilon^2}{c^2}\biggr)
-2\epsilon u'\gamma'},$$
where
$$\epsilon=u\cos\alpha-c^2(\gamma-1)\frac{\gamma'}{u'},$$
and $\alpha=\cos^{-1}(\u\cdot\u'/uu')$ is the angle between $\u$ and $\u'$.
Because the $u$ dependence only enters through $\epsilon$ and because
$\epsilon = O(u)$, we proceed by expanding Eq.~(\ref{pot-int-harm})
for small $\epsilon$.  We use Eq.~(\ref{gen-fun}) to expand
$y\subb 0k\s(w/c)$ as
\begin{equation}\label{gen-fun-a}
y\subb 0k\s(w/c)
=\sum_{l'=0}^\infty c^{-l'}y\subb{l'}k\s(u'/c)\frac{\epsilon^{l'}}{l'!}.
\end{equation}
When we substitute for $\epsilon$ and perform the angle integrations, we
encounter integrals of the form
$$I_{nl}=
 \int\!\frac{d\Omega}{4\pi}\,
 \cos^n\! \alpha\,Y_{lm}(\theta',\phi').
$$
Using formula (7.126.1) of Gradshteyn and Ryzhik \cite{GraRyz65}, we can expand
$\cos^n\!\alpha$ as
$$\cos^n\!\alpha=\sum_{k=0}^{\lfloor n/2\rfloor}\frac{n!}{2^k k!}
\frac{2n-4k+1}{(2n-2k+1)!!} P_{n-2k}(\cos\alpha).$$
Likewise, we can use Eq.~(3.62) of Jackson \cite{Jackso75} to expand
$P_n(\cos\alpha)$ as
$$P_n(\cos\alpha)=\frac{4\pi}{2n+1}\sum_{k=-n}^n Y^*_{nk}(\theta',\phi')
 Y_{nk}(\theta,\phi).$$
Substituting these series into $I_{nl}$ and using the orthogonality
condition for the spherical harmonics
$$ \int\!\! d\Omega\,
 Y^*_{l'm'}(\theta,\phi) Y_{lm}(\theta,\phi)=\delta_{l'l}\delta_{m'm},$$
we obtain
$$I_{nl}=
\begin{cases}
\displaystyle
  \frac{n!}{2^{(n-l)/2}((n-l)/2)!(n+l+1)!!},& for $n\ge l$ and $n-l$ even,\\
0, & otherwise.
\end{cases}
$$
Therefore, the first term
in the sum in Eq.~(\ref{gen-fun-a}) which contributes to the integral
in Eq.~(\ref{pot-int-harm}) is the term $l'=l$.  This results in a
term of order $u^l$.  The remaining terms in the sum contribute terms of
higher order in $u$.  Thus, we have
\begin{equation}\label{pot-int-exp}
\psi\subb{lm}k\s(u)\to\frac{u^lc^{2k-l-3}}{(2l+1)!!}
 \int_0^\infty y\subb lk\s(u'/c)\frac{u'^2}{\gamma'}f_{lm}(u')\,du'.
\end{equation}

Comparing Eqs.~(\ref{psi-lead}) and (\ref{pot-int-exp}), we find that
$C\subb{lm}k\s=0$, and therefore that $\psi^{\rm HS}\subb{lm}k\s\linebreak[0]=0$.
The desired solution is given just by the particular integral,
Eq.~(\ref{psi-pi}), which for $k>0$ we may write as
\begin{eqalignno}
\psi\subb{lm}k\s(u)&=
\int_0^u N\subb lk\s(u,u') \frac{u'^2}{\gamma'} f_{lm}(u')\,du'
\nonumber\\&\qquad{}+
\int_u^\infty N\subb lk\s(u',u) \frac{u'^2}{\gamma'} f_{lm}(u')\,du',
\label{general-sol}
\end{eqalignno}
with $N\subb lk\s$ given by Eqs.~(\ref{N-def}).
This completes the solution for the potentials $\psi\subb{lm}k\s$.

In obtaining this result, we have, in effect, found a spherical harmonic
decomposition of the kernel in Eq.~(\ref{pot-int})
\begin{eqalignno*}
c^{2k-3}y\subb 0k\s(w/c)
 &= 4\pi\sum_{l=0}^\infty\sum_{m=-l}^l N\subb lk\s(u_>,u_<)
  Y^*_{lm}(\theta',\phi')Y_{lm}(\theta,\phi)\\
&=\sum_{l=0}^\infty(2l+1)N\subb lk\s(u_>,u_<)P_l(\cos\alpha).
\end{eqalignno*}
For $k=1$, this yields an addition formula for a class of associated Legendre
functions:
$$
\frac{P^{1/2}_{a-1/2}(r)}{(r^2-1)^{1/4}}
= \sum_{l=0}^{\infty} (-1)^l(2l+1)\sqrt{\frac{\pi}2}\,
\frac{P^{l+1/2}_{a-1/2}(\gamma)}{(\gamma^2-1)^{1/4}}\,
\frac{P^{-l-1/2}_{a-1/2}(\gamma')}{(\gamma'^2-1)^{1/4}}\,
P_l(\cos\alpha),
$$
where $r=\gamma\gamma'-\sqrt{\gamma^2-1}\sqrt{\gamma'^2-1}\cos\alpha$ and
$1<\gamma'\le\gamma$. This identity is not found in the usual handbooks,
% such as Gradshteyn and Ryzhik \cite{GraRyz65} or Bateman \cite{Batema53},
although it did turn out to be known \cite{henric55,belooz76}.

\section{Nonrelativistic limit}\label{nonrelativistic}

Taking the limit $c\to\infty$ in the preceding equations, we recover the
well-known nonrelativistic collision operator. We catalog here the important
results.

The kernel $\UU$ reduces to the one given by Landau \cite{Landau36}
$$
\UU(\u,\u')=
\frac1{\abs{\u-\u'}^3}
\bigl({\textstyle\abs{\u-\u'}^2\II-(\u-\u')(\u-\u')}\bigr).
$$
The quantities $r$ and $w$ reduce to $1$ and $\abs{\u-\u'}$, respectively.
The potentials $\Psi\ssuba{s'}k\s$ only depend on the number of indices, $k$,
and not on their values. We will, therefore, drop the indices and write
$\Psi\ssuba{s'}k{}(\u)$ instead of $\Psi\ssuba{s'}k\s(\u)$. When
Eq.~(\ref{y-nr}) is substituted into Eq.~(\ref{pot-int}), the potentials
become
$$
\Psi\ssuba{s'}k{}(\u)=
\begin{cases}
  \displaystyle f_{s'}(\u),&for $k=0$,\\
  \displaystyle-\frac1{4\pi}\int\frac{|\u-\u'|^{2k-3}}{(2k-2)!}
    f_{s'}(\u')\,d^3\u',&for $k>0$.
\end{cases}
$$
The potentials satisfy $L\Psi\ssuba{s'}{k+1}{}=\Psi\ssuba{s'}k{}$, where $L$ is
now the velocity-space Laplacian,
$$
L\Psi=\frac{\d}{\d\u}\cdot\frac{\d\Psi}{\d\u}.
$$
In particular, we have
\begin{eqalignno*}
L\Psi\ssuba{s'}1{}&=f_{s'},\\
L\Psi\ssuba{s'}2{}&=\Psi\ssuba{s'}1{}.
\end{eqalignno*}
The diffusion and friction coefficients are given by
\begin{eqalignno*}
\DD^{s/s'}(\u)&=-\frac{4\pi\Gamma^{s/s'}}{n_{s'}}\,
  \frac{\d^2}{\d\u\d\u}\Psi\ssuba{s'}2{},\\
\F^{s/s'}(\u)&=-\frac{4\pi\Gamma^{s/s'}}{n_{s'}}\frac{m_s}{m_{s'}}\,
  \frac{\d}{\d\u} \Psi\ssuba{s'}1{}.
\end{eqalignno*}

The homogeneous solutions to the separated radial components of the
nonrelativistic potential equations are given by Eqs.~(\ref{jy-nr}).
Substituting these into Eqs.~(\ref{N-def}) and then substituting the
result into Eq.~(\ref{general-sol}), we obtain the Legendre harmonic
expansion for the potentials
\begin{eqalignno*}
&\psi\subb{lm}1{}(u)=\\*
&\quad
-\int_0^u \frac1{2l+1}\frac{u'^l}{u^{l+1}} u'^2 f_{lm}(u')\,du'
-\int_u^\infty \frac1{2l+1}\frac{u^l}{u'^{l+1}} u'^2 f_{lm}(u')\,du',\quad\\
&\psi\subb{lm}2{}(u)=\\*
&\quad
-\int_0^u \frac1{2(2l+1)}\biggl[\frac1{2l+3}\frac{u'^{l+2}}{u^{l+1}}
-\frac1{2l-1}\frac{u'^{l}}{u^{l-1}}\biggr] u'^2 f_{lm}(u')\,du'\\*
&\quad-\int_u^\infty\frac1{2(2l+1)}\biggl[\frac1{2l+3}\frac{u^{l+2}}{u'^{l+1}}
-\frac1{2l-1}\frac{u^{l}}{u'^{l-1}}\biggr] u'^2 f_{lm}(u')\,du'.
\end{eqalignno*}
These expressions coincide with the results of Rosenbluth, MacDonald, and
Judd \cite{RMJ57}.

\section{Isotropic background}\label{isotropic}

In this section and in Sec.~\ref{first-harmonic}, we consider the cases
where the background is described by the $l=0$, $m=0$ and $l=1$, $m=0$
components in Eq.~(\ref{spher-harm}).  First let us consider the
azimuthally symmetric case, $m=0$ and $l$ arbitrary.  For simplicity we
will drop the $m$ subscript and thus write
$$
\Psi\suba k\s(\u)=\sum_{l=0}^\infty \psi\subb lk\s(u)P_l(\cos\theta).
$$
When we substitute this representation into Eqs.~(\ref{lk-comp}), the
second derivatives in the component $\LL_{uu}\Psi$ may be eliminated by using
the differential equation~(\ref{spher-eq}); this gives
\begin{eqalignno}\label{Luu-comp}
&\LL_{uu}\bigl(\psi\subb l{k+1}{\s a}(u)P_l(\cos\theta)\bigr) =
\biggl[ \gamma^2\psi\subb lk\s
  - \frac{2\gamma^4}{u}\pd{\psi\subb l{k+1}{\s a}}u\nonumber\\
&\qquad\qquad{}+\gamma^2\biggl(\frac{l(l+1)}{u^2}
  + \frac{a^2-1}{c^2}\biggr)\psi\subb l{k+1}{\s a}
\biggr]\,P_l(\cos\theta).
\end{eqalignno}

Let us now specialize to an isotropic background $l=0$, $P_0(\cos\theta)=1$.
If we substitute Eqs.~(\ref{lk-comp}) and~(\ref{Luu-comp}) into
Eqs.~(\ref{diff-fric}), we obtain
\begin{subequations}
\begin{eqalignno}
D_{uu,0}^{s/s'}&=
  \frac{4\pi\Gamma^{s/s'}}{n_{s'}}\frac\gamma{u}
  \biggl[2\gamma^2\frac{d\psi\ssubb{s'}02{02}}{du}-u\psi\ssubb{s'}01{0}
\nonumber\\*
&\qquad\qquad\qquad{}
  -\frac{8\gamma^2}{c^2}\frac{d\psi\ssubb{s'}03{022}}{du}
  +\frac{8u}{c^4}\psi\ssubb{s'}03{022}\biggr],\\
D_{\theta\theta,0}^{s/s'}&=
  \frac{4\pi\Gamma^{s/s'}}{n_{s'}}\frac1{\gamma u}
  \biggl[-\gamma^2\frac{d\psi\ssubb{s'}02{02}}{du}
  -\frac u{c^2}\psi\ssubb{s'}02{02}
\nonumber\\*
&\qquad\qquad\qquad{}
  +\frac{4\gamma^2}{c^2}\frac{d\psi\ssubb{s'}03{022}}{du}
  -\frac{4u}{c^4}\psi\ssubb{s'}03{022}\biggr],\\
F_{u,0}^{s/s'}&=
  \frac{4\pi\Gamma^{s/s'}}{n_{s'}}\frac{m_s}{m_{s'}}\gamma
  \biggl[-\frac{d\psi\ssubb{s'}01{1}}{du}
  +\frac2{c^2}\frac{d\psi\ssubb{s'}02{11}}{du}\biggr].
\end{eqalignno}
\end{subequations}
The other components of $\DD_0^{s/s'}$ and $\F_0^{s/s'}$ vanish.
Lastly we substitute for the potentials using
Eq.~(\ref{general-sol}).  Some massaging of the result leads to
\begin{subequations}\label{diff-iso}
\begin{eqalignno*}
D_{uu,0}^{s/s'}&=
\frac{4\pi\Gamma^{s/s'}}{n_{s'}}\biggl\{
\int_0^u \bigl[2\gamma^2c^2 j'\subb 02{02} - 8c^2 j'\subb 03{022} \bigr]
\frac{\gamma}{u^3} \frac{u'^2}{\gamma'} f_{s'0}(u')\,du'\\
&\qquad{}+\int_u^\infty
\bigl[2\gamma'^2c^2 j\subb 02{02} - 8c^2 j\subb 03{022}\bigr]
\frac{\gamma}{u^2} \frac{u'}{\gamma'} f_{s'0}(u') \,du'\biggr\},\yesnumber\\
D_{\theta\theta,0}^{s/s'}&=
\frac{4\pi\Gamma^{s/s'}}{n_{s'}}\biggl\{
\int_0^u \biggl[\frac12 j'\subb 012
  - \biggl(\frac{c^2}{u^2}+\frac1{\gamma^2}\biggr)
     j'\subb 02{02}\\[-4pt]
&\qquad\qquad\qquad\qquad{}
+\frac{4}{\gamma^2}\frac{c^2}{u^2}
   j'\subb 03{022}\biggr]
\frac{\gamma}u \frac{u'^2}{\gamma'} f_{s'0}(u') \,du'\\
&\qquad{}+\int_u^\infty
\biggl[\frac12 \frac{\gamma'^2}{\gamma^2}j\subb 012
  - \frac{u'^2}{u^2} \biggl(\frac{c^2}{u'^2}+\frac1{\gamma^2}\biggr)
     j\subb 02{02}\\[-4pt]
&\qquad\qquad\qquad\qquad{}
+\frac4{\gamma^2}\frac{c^2}{u^2} j\subb 03{022}\biggr]
\gamma \frac{u'}{\gamma'} f_{s'0}(u') \,du'\biggr\},\yesnumber\\
F_{u,0}^{s/s'}&=
-\frac{4\pi\Gamma^{s/s'}}{n_{s'}}\frac{m_s}{m_{s'}}\biggl\{
\int_0^u \bigl[\gamma^2 j'\subb 011 - 2 j'\subb 02{11}\bigr]
\frac 1{u^2}\frac{u'^2}{\gamma'} f_{s'0}(u') \,du'\\*
&\qquad{}+\int_u^\infty 4\frac{u'}u j\subb 02{02}
f_{s'0}(u') \,du'\biggr\},\yesnumber
\end{eqalignno*}
\end{subequations}
where $j\subb lk\s=j\subb lk\s(u/c)$ and $j'\subb lk\s=j\subb lk\s(u'/c)$.

Of particular interest is the case of a Maxwellian background, i.e.,
$$f_{s'0}(u)=f_{s'\maxw}=
\frac{n_{s'} m_{s'}}{4\pi c T_{s'} K_2(m_{s'}c^2/T_{s'})}
\exp\biggl(-\frac{m_{s'}c^2\gamma}{T_{s'}}\biggr),$$
where $K_n$ is the $n$th order Bessel function of the second kind.  First
of all we can verify that
$$F_{u,0}^{s/s'}=-\frac{m_s v}{T_{s'}}D_{uu,0}^{s/s'},$$
where $v=u/\gamma$.  This is accomplished by substituting
$$f_{s'\maxw}(u')=-\frac{T_{s'}}{m_{s'}v'}\frac d{du'}f_{s'\maxw}(u')$$
into the expression for $D_{uu,0}^{s/s'}$ and integrating by parts.  This
relation between $F_{u,0}^{s/s'}$ and $D_{uu,0}^{s/s'}$ implies that the
collisions will cause $f_s$ to relax to a Maxwellian with temperature $T_{s'}$.

In the high-energy limit $m_{s'}c^2(\gamma-1)\gg T_{s'}$, the indefinite limits
in the integrals in Eqs.~(\ref{diff-iso}) can be replaced by $\infty$.
We can perform the resulting integrals using formula (7.141.5) of Gradshteyn
and Ryzhik \cite{GraRyz65} which gives, after a change of integration variable,
$$
\int_0^\infty \exp\bigl(-\beta\sqrt{1+z^2}\bigr)
\frac{z^{l+2}}{\gamma} j\subb l1a(z)\,dz=
\frac{K_a(\beta)}{\beta^{l+1}}.
$$
For $l=0$ and $a=2$ this gives the normalization condition for the
Maxwellian
$$\int_0^\infty 4\pi u^2 f_{s'\maxw}(u) \,du = n_{s'}.$$
On carrying out the integrations in Eqs.~(\ref{diff-iso}), we obtain
\begin{subequations}
\begin{eqalignno}
D_{uu,0}^{s/s'}&=\Gamma^{s/s'}
\frac{u_{ts'}^2}{v^3}\frac{K_1}{K_2}
\biggl(1-\frac{K_0}{K_1} \frac{u_{ts'}^2}{\gamma^2c^2}\biggr),\\
D_{\theta\theta,0}^{s/s'}&=\Gamma^{s/s'}\frac1{2 v}\biggl[1-\frac{K_1}{K_2}
\biggl(\frac{u_{ts'}^2}{u^2}+\frac{u_{ts'}^2}{\gamma^2c^2}\biggr)+
\frac{K_0}{K_2}\frac{u_{ts'}^2}{u^2}\frac{u_{ts'}^2}{\gamma^2c^2}\biggr],\\
F_{u,0}^{s/s'}&=-\Gamma^{s/s'}\frac{m_s}{m_{s'}}\frac1{v^2}
\frac{K_1}{K_2}
\biggl(1-\frac{K_0}{K_1} \frac{u_{ts'}^2}{\gamma^2c^2}\biggr),
\end{eqalignno}
\end{subequations}
where $u_{ts}^2=T_s/m_s$ and the argument for the Bessel functions is
$m_{s'}c^2/T_{s'}$.
In the limit $m_{s'}\to\infty$, we recover the Lorentz collision operator
\begin{subequations}\label{lorentz}
\begin{eqalignno}
D_{uu,0}&=F_{u,0}=0,\\
D_{\theta\theta,0}&=\Gamma^{s/s'}\frac1{2 v}.
\end{eqalignno}
\end{subequations}

\section{First harmonic}\label{first-harmonic}

In addition to the isotropic components of the collision operator
calculated in the previous section, the first harmonic of the Legendre
expansion is also required in the calculation of the electrical
conductivity.  Specifically, we need to compute the term
$C^{s/s'}(f_{s\maxw},f_{s'1}\cos\theta)$.
We can express this in terms of the potentials and their
derivatives using Eqs.~(\ref{coll-op}), (\ref{diff-fric}), and
(\ref{lk-comp}) to give
\begin{eqalignno*}
\hspace{2em}&\hspace{-2em}
\frac{C^{s/s'}(f_{s\maxw}(u),f_{s'1}(u)\cos\theta)}{f_{s\maxw}(u)\cos\theta}\\
&=\frac{4\pi\Gamma^{s/s'}}{n_{s'}}\biggl\{
\frac{m_s}{m_{s'}} \biggl[\frac1\gamma \psi\ssubb{s'}10{}
- \frac u{u_{ts}^2} \frac{d\psi\ssubb{s'}11{1}}{du}
\\&\qquad\qquad\qquad\qquad{}
- \frac2{c^2 \gamma} \psi\ssubb{s'}11{1}
+ \frac{2u}{c^2u_{ts}^2} \frac{d\psi\ssubb{s'}12{11}}{du}\biggr]\\
&\qquad\qquad{}
+ \frac u{u_{ts}^2} \frac{d\psi\ssubb{s'}11{0}}{du}
- \biggl(\frac{u^2}{\gamma u_{ts}^4} - \frac1{u_{ts}^2}\biggr) \psi\ssubb{s'}11{0}\\
&\qquad\qquad{}
+\biggl(\frac{2\gamma u}{u_{ts}^4} - \frac {2u}{c^2u_{ts}^2}\biggr)
\frac{d\psi\ssubb{s'}12{02}}{du}
-\biggl(\frac2{\gamma u_{ts}^4}-\frac2{c^2u_{ts}^2}\biggr)
\psi\ssubb{s'}12{02}\\
&\qquad\qquad{}
-\frac {8\gamma u}{c^2u_{ts}^4} \frac{d\psi\ssubb{s'}13{022}}{du}
+\frac{8\gamma}{c^2u_{ts}^4}\psi\ssubb{s'}13{022}\biggr\}\yesnumber.
\end{eqalignno*}
Finally, we substitute for the potentials and simplify to obtain
\begin{eqalignno*}
\hspace{2em}&\hspace{-2em}
\frac{C^{s/s'}(f_{s\maxw}(u),f_{s'1}(u)\cos\theta)}{f_{s\maxw}(u)\cos\theta}
\\*
&=\frac{4\pi\Gamma^{s/s'}}{n_{s'}}\biggl\{
\frac{m_s}{m_{s'}}\frac1\gamma f_{s'1}(u)
\\&\qquad{}+
\int_0^u\biggl[
\frac1{u^2}\biggl(
  2\frac{m_s}{m_{s'}}\frac{j'\subb11{1}}{c^2}
  +\frac{j'\subb11{2}}{u_{ts}^2}
  -10\frac{j'\subb12{02}}{u_{ts}^2}\biggr)\\*
&\qquad\qquad{}+\frac\gamma{u^2}\biggl(
  -2\frac{m_s}{m_{s'}}\frac{j'\subb11{1}}{u_{ts}^2}
  +4\frac{m_s}{m_{s'}}\frac{j'\subb12{11}}{u_{ts}^2}\\[-4pt]
&\qquad\qquad\qquad\qquad\qquad{}
  +6\frac{c^2j'\subb12{02}}{u_{ts}^4}
  -24\frac{c^2j'\subb13{022}}{u_{ts}^4}\biggr)\\
&\qquad\qquad{}+\biggl(\frac{j'\subb11{0}}{c^2u_{ts}^2}\biggr)
  +\gamma\biggl(2\frac{j'\subb12{02}}{u_{ts}^4}\biggr)\biggr]
\frac{cu'^2}{\gamma\gamma'}f_{s'1}(u')\,du'\\
&\qquad{}+
\int_u^\infty\biggl[
\frac1{u'^2}\biggl(
  2\frac{m_s}{m_{s'}}\frac{j\subb11{1}}{c^2}
  +\frac{m_s}{m_{s'}}\frac{j\subb11{2}}{u_{ts}^2}
  -10\frac{m_s}{m_{s'}}\frac{j\subb12{02}}{u_{ts}^2}\biggr)\\
&\qquad\qquad{}+\frac{\gamma'}{u'^2}\biggl(
  -2\frac{j\subb11{1}}{u_{ts}^2}
  +4\frac{j\subb12{11}}{u_{ts}^2}\\[-4pt]
&\qquad\qquad\qquad\qquad\qquad{}+
  6\frac{c^2j\subb12{02}}{u_{ts}^4}
  -24\frac{c^2j\subb13{022}}{u_{ts}^4}\biggr)\\
&\qquad\qquad{}+\biggl(
  \frac{m_s}{m_{s'}}\frac{j\subb11{0}}{c^2u_{ts}^2} \biggr)
  +\gamma'\biggl(2\frac{j\subb12{02}}{u_{ts}^4}\biggr)\biggl]
\frac{cu'^2}{\gamma\gamma'}f_{s'1}(u')\,du'\biggr\},
\yesnumber\label{first-harm}
\end{eqalignno*}
where, as before, $j\subb lk\s=j\subb lk\s(u/c)$ and $j'\subb lk\s=j\subb
lk\s(u'/c)$.

Both in Eqs.~(\ref{diff-iso}) and in Eq.~(\ref{first-harm}), one
can substitute for $j\subb lk\s$ from Eqs.~(\ref{l=0}) and~(\ref{l=1})
and thereby express the collision operator entirely in terms of elementary
functions.  The resulting expressions would be very badly behaved numerically
near $u=0$ because of large cancellations.  It is preferable, therefore, to
evaluate $j\subb lk\s$ directly by the method outlined in the appendix.

The collision operator obeys the conservation law
$$
\int \bigl[h_s C^{s/s'}(f_s, f_{s'})
         + h_{s'} C^{s'/s}(f_{s'}, f_s)\bigr]\,d^3\u=0,
$$
where $h_s=a_0 + {\bf a}_1\cdot m_s \u + a_2 m_s c^2 \gamma$, and $a_0$,
${\bf a}_1$, and $a_2$ are arbitrary constants.
The collision operator is also self-adjoint:
$$
\int\psi C^{s/s'}(\chi f_{s\maxw},f_{s'\maxw})\,d^3\u=
  \int\chi C^{s/s'}(\psi f_{s\maxw},f_{s'\maxw})\,d^3\u,
$$
and satisfies the symmetry
$$
\int\psi C^{s/s'}(f_{s\maxw},\chi f_{s'\maxw})\,d^3\u=
  \int\chi C^{s'/s}(f_{s'\maxw},\psi f_{s\maxw})\,d^3\u,
$$
where $\psi$ and $\chi$ are arbitrary functions of $\u$, and $T_s=T_{s'}$.
Combining these two properties gives
$$
C^{s/s'}(h_s f_{s\maxw}, f_{s'\maxw}) +
  C^{s/s'}(f_{s\maxw}, h_{s'} f_{s'\maxw}) = 0.
$$
This provides a useful check on the implementations of
Eqs.~(\ref{diff-iso}) and (\ref{first-harm})

\section{Calculation of the conductivity}\label{conductivity}

At this point the calculation of the electrical conductivity is
straightforward.  We consider an electron-ion plasma with infinitely
massive stationary ions; $m_i\to\infty$ and $f_{i\maxw}\to n_i\delta(\u)$.
In the presence of a weak electric field $E\hat{\bf z}$, the electron
distribution is given to first order by $f_{e\maxw}
(1+\chi_1(u,t)\cos\theta)$.  The linearized Boltzmann equation may be
written in the form
\begin{equation}\label{spitz-eq}
\frac{\d\chi_1}{\d t}=\frac{q_eEv}{T_e}+\hat C_e(\chi_1),
\end{equation}
where $\hat C_e(\chi_1)$ is the linearized electron collision term
\begin{eqalignno*}
\hat C_e(\chi_1)
&= \frac{1}{f_{e\maxw}\cos\theta}
\bigl[C^{e/e}(f_{e\maxw}\chi_1\cos\theta,f_{e\maxw})
+ C^{e/e}(f_{e\maxw},f_{e\maxw}\chi_1\cos\theta)\\&\qquad\qquad\qquad\qquad{}
+ C^{e/i}(f_{e\maxw}\chi_1\cos\theta,f_{i\maxw})\bigr].
\end{eqalignno*}
The first term here is given by
$$\frac{C^{e/e}(f_{e\maxw}\chi_1\cos\theta,f_{e\maxw})}{f_{e\maxw}\cos\theta}
= \frac1{u^2}\frac{\d}{\d u} u^2 D_{uu,0}^{e/e} \frac{\d\chi_1}{\d u}
+ F_{u,0}^{e/e}\frac{\d\chi_1}{\d u}
- \frac2{u^2}D_{\theta\theta,0}^{e/e}\chi_1,$$
with $\DD$ and $\F$ given by Eqs.~(\ref{diff-iso}).
The second is given directly by Eq.~(\ref{first-harm}) with $s=s'=e$
and $f_{s'1}=f_{s'\maxw}\chi_1$.  The last term is given by the Lorentz limit,
Eqs.~(\ref{lorentz}),
$$\frac{C^{e/i}(f_{e\maxw}\chi_1\cos\theta,f_{i\maxw})}{f_{e\maxw}\cos\theta}
=-\frac{\Gamma^{e/i}}{u^2v}\chi_1.$$

The conductivity is defined by
$$
\sigma=\frac{4\pi q_e}{3E}\int_0^\infty f_{em}(u)\chi_1(u,t\to\infty)vu^2\,du.
$$
The time asymptotic solution $\chi_1(u,t\to\infty)$ is determined by solving
Eq.~(\ref{spitz-eq}) as an initial value problem.  We take
$\chi_1(u,t=0)=0$ (for example).  The differential terms
$C^{e/e}(f_{e\maxw}\chi_1\cos\theta,f_{e\maxw})$ and
$C^{e/i}(f_{e\maxw}\chi_1\cos\theta,f_{i\maxw})$ are both treated fully
implicitly, while the integral term
$C^{e/e}(f_{e\maxw},f_{e\maxw}\chi_1\cos\theta)$ is treated explicitly.
This permits large time steps to be taken and leads to a rapid convergence
to a steady state.  A check on $\chi_1$ is obtained by evaluating the first
moment of the linearized Boltzmann equation.  The electron-electron collision
terms drop out by conservation of momentum, to give
\begin{equation}\label{mom-balance}
\frac{4\pi}3\int_0^\infty f_{em}\chi_1\gamma\,du=
\frac{n_e q_e E}{m_e Z \Gamma^{e/e}},
\end{equation}
where
$$
Z=\frac{\Gamma^{e/i}}{\Gamma^{e/e}}
=-\frac{q_i\log\Lambda^{e/i}}{q_e\log\Lambda^{e/e}}
$$
is the effective ion charge state, and where we have assumed $q_en_e+q_in_i=0$.

It is convenient to write $\sigma$ as
\begin{eqalignno*}
\sigma&=\frac{n_e q_e^2 u_{te}^3}{m_e Z\Gamma^{e/e}}
\sigman(\Theta,Z)\\
&=\frac{4\pi\epsilon_0^2}{m_e^{1/2}q_e^2\log\Lambda^{e/e}}\frac{T_e^{3/2}}{Z}
\sigman(\Theta,Z),
\end{eqalignno*}
where
$$
\Theta=\frac{T_e}{m_ec^2}=\frac{T_e}{511\,\rm keV},\qquad u_{te}=\sqrt{T_e/m_e}.
$$
The normalized conductivity $\sigman$ is a dimensionless function of two
dimensionless arguments.  In the limit $\Theta\to0$, $\sigman$ is bounded
and nonzero, and we recover the nonrelativistic scaling $\sigma\propto
T_e^{3/2}$. Values of $\sigman$ for various $\Theta$ and $Z$ are tabulated in
Table~1 and plotted in Fig.~\ref{fig-1}.  The nonrelativistic conductivity
was first calculated by Spitzer and H\"arm \cite{SpiHar53}, who quote values
of $\sigman(0,Z)/\sigman(0,\infty)$.  Their results coincide with ours in the
limit $\Theta\to0$.

In the limit $Z\to\infty$, electron-electron collisions can be ignored,
and the relevant collision term is the Lorentz electron-ion collision term,
Eqs.~(\ref{lorentz}).  This case is considered by Lifshitz and Pitaevskii
\cite{LifPit81}.  We can write
$$\chi_1=\frac{q_e E}{ZT_e\Gamma^{e/e}} u^2v^2.$$
The resulting conductivity is
\begin{eqalignno*}
\sigman&=\frac{4\pi}3\frac1{n_eu_{te}^5}
\int_0^\infty u^4v^3f_{em}\,du\\
&=\frac1{3\Theta^{7/2}K_2(\Theta^{-1})}
\int_1^\infty\frac{(\gamma^2-1)^3}{\gamma^2}
\exp\biggl(-\frac\gamma\Theta\biggr)\,d\gamma.\\
\end{eqalignno*}
Evaluating the integral, we obtain
\begin{eqalignno}
\sigman&=\frac1{3\Theta^{7/2}K_2(\Theta^{-1})}
  \biggl(\frac{E_1(\Theta^{-1})}\Theta\nonumber\\
&\qquad\qquad{}-(1-\Theta+2\Theta^2-6\Theta^3-24\Theta^4-24\Theta^5)
  \exp(\Theta^{-1})\biggr), \label{Z-inf}
\end{eqalignno}
where $E_n$ is the exponential integral.  In the limit $\Theta\to0$, this
reduces to $\sigman=16\sqrt{2/\pi}$.  For $\Theta\to\infty$, we obtain
$\sigman=4/\sqrt{\Theta}$, which agrees with the result of Lifshitz and
Pitaevskii.

Another tractable, albeit less interesting, limit is $Z\to0$.  In this case,
this electrons equilibrate with themselves so that their distribution is a
Maxwellian drifting at $v_d$ and $\chi_1 = v_d u/u_{te}^2$.  The drift
speed $v_d$ is found by applying Eq.~(\ref{mom-balance}) to give
$$
v_d=\frac{3\exp(\Theta^{-1})K_2(\Theta^{-1})}
{\sqrt{\Theta}(1+2\Theta+2\Theta^2)}
\frac{q_eE u_{te}^3}{m_e Z \Gamma^{e/e}}.
$$
The resulting conductivity is
\begin{equation} \label{Z-0}
\sigman = \frac{3\exp(\Theta^{-1})K_2(\Theta^{-1})}
{\sqrt{\Theta}(1+2\Theta+2\Theta^2)}.
\end{equation}
In this expression we recognize the result obtained by van Erkelens and
van Leeu\-wen \cite{ErkLee84} on the basis of a lowest-order variational
treatment of the relativistic Boltzmann equation. Their result for the
conductivity of a relativistic plasma, therefore, corresponds to the limit
$Z\to0$. For this case the limit $\Theta\to0$ gives $\sigman=3\sqrt{\pi/2}$
and the limit $\Theta\to\infty$ gives $\sigman=3/\sqrt{\Theta}$.

\section{Conclusions}

In our earlier work \cite{BraKar87}, we gave a differential formulation for
the collision operator for a relativistic plasma.  This
formulation is summarized by Eqs.~(\ref{pot-diff-prl}) and
(\ref{diff-fric}).  A major objective of the present work is to solve the
differential equations~(\ref{pot-diff-prl}) and, hence, to express the
potentials in terms of quadrature.  This was achieved by using an expansion
in Legendre harmonics, Eq.~(\ref{spher-harm}).  The radial
components of the potentials are then given by Eq.~(\ref{general-sol})
where the kernels $N\subb lk\s(u,u')$ are given by Eqs.~(\ref{N-def});
these in turn involve the special functions $j\subb lk\s$ and $y\subb
lk\s$, whose properties are given in the appendix.  The entire formulation
is well-behaved in the nonrelativistic limit; indeed, in this limit, the
potentials and their solutions agree with the earlier nonrelativistic
treatment of Rosenbluth {\it et al.} \cite{RMJ57} and Trubnikov
\cite{Trubni58}.

Several computer codes exist which solve the nonlinear
Fokker-Planck equation in the nonrelativistic limit.  In many of these
codes the collision operator is evaluated in terms of a Legendre harmonic
expansion of the potentials.  Our results are easily incorporated into such
codes, allowing them to treat relativistic collisions.  The evaluation of
the collision operator will be a few times more costly than in the
nonrelativistic case, firstly because five potentials need to be computed
instead of two, and secondly because the kernels involve the special
functions $j\subb lk\s$ and $y\subb lk\s$ instead of simple powers of $u$.

As an application of this formulation, we give in Eqs.~(\ref{diff-iso})
explicit forms for the diffusion and friction coefficients for an isotropic
background.  Finally, we calculate the electrical conductivity of an
electron-ion plasma with massive ions.  Our results agree with those of
Spitzer and H\"arm \cite{SpiHar53} in the nonrelativistic limit.  We also
give analytical expressions for the conductivity for the limiting cases $Z
\to\infty$ and $Z\to 0$ in Eqs.~(\ref{Z-inf}) and (\ref{Z-0}).

\section*{Acknowledgement}

This work was supported by the U.S. Department of Energy under contract
DE--AC02--76--CHO--3073.

\appendix
\section{Properties of the homogeneous solutions}

In this appendix we will develop some properties of the solutions of the
homogeneous radial equations (\ref{homo}).

\subsection{Definitions}

First we shall obtain fundamental solutions $j\subb l1a(z)$ and $y\subb l1a(z)$
to the lowest-order homogeneous differential equation
\begin{equation}\label{homo-1}
  L_{l,a}\chi(z)=0,
\end{equation}
where
\begin{equation}\label{Ll-def}
L_{l,a}\chi(z)=(1+z^2)\frac{d^2\chi}{dz^2}
  +\biggl(\frac 2z+3z\biggr)\frac{d\chi}{dz}
  -\biggl(\frac{l(l+1)}{z^2}+a^2-1\biggr)\chi.
\end{equation}
The variable $z$ corresponds to $u/c$ in the main text. In this appendix, $l$
must be an integer, $a$ must be real, and $z$ is in the complex plane cut along
the negative real axis.

By changing the independent variable to $\gamma=\sqrt{1+z^2}$ and defining a
new dependent variable $\xi(\gamma)=\sqrt{z}\chi(z)$, we obtain the equation
$$
(\gamma^2-1)\frac{d^2\xi}{d\gamma^2} + 2\gamma\frac{d\xi}{d\gamma}
  -\biggl[{\textstyle(a+\frac12)(a-\frac12)}
    +\frac{(l+\frac12)^2}{\gamma^2-1}\biggr]\xi = 0.
$$
This is the generalized Legendre equation, whose solutions are the associated
Legendre functions $P^{-l-1/2}_{a-1/2}(\gamma)$ and
$P^{l+1/2}_{a-1/2}(\gamma)$.
We choose to define $j\subb l1a$ and $y\subb l1a$ by
\begin{eqalignno}
j\subb l1a(z)\label{A-j-def}
  &=\sqrt{\frac{\pi}{2z}}P^{-l-1/2}_{a-1/2}(\gamma),\\
y\subb l1a(z)\label{A-y-def}
  &=(-1)^{l+1}\sqrt{\frac{\pi}{2z}}P^{l+1/2}_{a-1/2}(\gamma).
\end{eqalignno}
An alternative representation is obtained by Whipple's transformation (see
Eq.\ (8.739) of Gradshteyn and Ryzhik \cite{GraRyz65})
\begin{eqalignno*}
j\subb l1a(z)
  &=\frac1z\frac{e^{-ia\pi}}{\Gamma(a+l+1)}Q^a_l(\gamma/z),\\
y\subb l1a(z)
  &=\frac1z\frac{(-1)^le^{-ia\pi}}{\Gamma(a-l)}Q^a_{-l-1}(\gamma/z).
\end{eqalignno*}
The Wronskian of the pair $j\subb l1a$ and $y\subb l1a$ follows from
Eq.~(8.741) of Gradshteyn and Ryzhik \cite{GraRyz65}; for integer $l$ and
arbitrary $a$ it is given by
\begin{equation}\label{wronskian}
j\subb l1a\frac d{dz}y\subb l1a - y\subb l1a\frac d{dz}j\subb l1a
  = \frac{1}{\gamma z^2}.
\end{equation}
Therefore, the solutions $j\subb l1\s$ and $y\subb l1\s$ are independent. When
$l$ is a non-negative integer, then $j\subb l1\s$ is regular at the origin while
$y\subb l1\s$ is singular.

In order to express the solutions to the higher-order equations, we
introduce the functions $j\subb lk\s(z)$ and $y\subb lk\s(z)$ where the
subscript ``$\s$'' stands for a string of $k$ indices.  The functions $j\subb
lk\s$ are defined recursively by
\begin{subequations}\label{high-def}
\begin{equation}
j\subb l{k+2}{\s aa'}(z) =
\frac{j\subb l{k+1}{\s a}(z)-j\subb l{k+1}{\s a'}(z)}{a^2-a'^2};
\end{equation}
the case $a=a'$ is handled by taking the limit
\begin{equation}
j\subb l{k+2}{\s aa}(z)=\pd{j\subb l{k+1}{\s a}(z)}{(a^2)}.
\end{equation}
\end{subequations}
The functions $y\subb lk\s(z)$ are defined in the same way, replacing
$j$ by $y$ throughout.  Making use of the property $(L_{l,a}-L_{l,a'})\chi
= -(a^2-a'^2)\chi$, it is readily established that $j\subb lk\s$ and
$y\subb lk\s$ satisfy
\begin{subequations}\label{homo-k}
\begin{eqalignno}
L_{l,a}j\subb l{k+1}{\s a}(z)&=j\subb lk\s(z),\\
L_{l,a}y\subb l{k+1}{\s a}(z)&=y\subb lk\s(z).
\end{eqalignno}
\end{subequations}

The functions $j\subb lk\s$ and $y\subb lk\s$ are invariant under permutation
of the indices in $\s$ and invariant under change of sign of any index in $\s$;
these properties reflect the commutivity of $L_{l,a}$ and $L_{l,a'}$ and
the symmetry $L_{l,a}=L_{l,-a}$. Also, we can write
\begin{equation}\label{y-j}
y\subb lk{\s}=(-1)^{l+1}j\subb {-l-1}k{\s},
\end{equation}
which reflects $L_{l,a}=L_{-l-1,a}$; and finally we have
$j\subb lk\s(-z) = (-1)^lj\subb lk\s(z)$ and
$y\subb lk\s(-z) = (-1)^ly\subb lk\s(z)$.
In most of what follows, we list only the properties of $j\subb lk\s$.
Equation~(\ref{y-j}) may be used to give the corresponding properties of
$y\subb lk\s$.

\subsection{Taylor series}

In the limit $z\to0$, we can expand $P^{-l-1/2}_{a-1/2}(\gamma)$ in a
Taylor series using formula (3.2.20) of Bateman \cite{Batema53}
$$
P^\mu_\nu(\gamma)=\frac{2^\mu(\gamma^2-1)^{-\mu/2}}{\Gamma(1-\mu)}
  {\textstyle F(\frac12+\frac12\nu-\frac12\mu,-\frac12\nu-\frac12\mu;
    1-\mu;1-\gamma^2)},
$$
where $F$ is the hypergeometric function
$$
F(a,b;c;z)=1+\frac{ab}{c}\frac{z}{1!}
  +\frac{a(a+1)b(b+1)}{c(c+1)}\frac{z^2}{2!}+\ldots.
$$
Substituting this into Eq.~(\ref{A-j-def}), we obtain a Taylor-series
expansion of $j\subb l1a(z)$
\begin{equation}\label{taylor-1}
j\subb l1a(z)=\frac{z^l}{(2l+1)!!}
  F\biggl(\frac{l+1-a}2,\frac{l+1+a}2;l+\frac32;-z^2\biggr),
\end{equation}
where $n!!$ denotes a double factorial
$$
n!!=
\begin{cases}
  \displaystyle 2^{n/2}\Gamma(\sfrac12n+1),&for $n$ non-negative and even,\\
  \displaystyle \sqrt{\frac2\pi}2^{n/2}\Gamma(\sfrac12n+1),&for $n$ odd.
\end{cases}
$$
With this definition, we find $(2n-1)!!(-2n-1)!!=(-1)^n$.

Another useful Taylor series is obtained by a transformation of the
hypergeometric function (Eq.~(9.131) of Gradshteyn and Ryzhik \cite{GraRyz65})
$$
F(a,b;c;z)=(1-z)^{c-a-b}F(c-a,c-b;c;z).
$$
This gives
\begin{equation}\label{taylor-2}
\frac{j\subb l1a(z)}{\gamma} = \frac{z^l}{(2l+1)!!}
F\biggl(\frac{l+2-a}2,\frac{l+2+a}2;l+\frac32;-z^2\biggr).
\end{equation}
Series~(\ref{taylor-1}) and (\ref{taylor-2}) either terminate (i.e.,
converge everywhere) or converge inside the circle $|z|=1$. In particular,
if $a$ is an integer and $l<\abs a$, then one of the series will terminate.

It is not so easy to write down Taylor series for the higher order functions
$j\subb lk\s(z)$ ($k>1$). However, the leading term is independent of the
indices making up $\s$, and can be obtained by substituting a series solution
into Eqs.~(\ref{homo-k}). One finds
\begin{subequations}\label{leading-order}
\begin{eqalignno}
j\subb lk\s(z)&=\frac{z^{l+2k-2}}{(2k-2)!!(2l+2k-1)!!}+O(z^{l+2k}),
  \label{leading-j}\\
y\subb lk\s(z)&=\frac{(-1)^k(2l-2k+1)!!}{(2k-2)!!z^{l-2k+3}}+O(1/z^{l-2k+1}).
  \label{leading-y}
\end{eqalignno}
\end{subequations}

\subsection{Asymptotic series}

Since $j\subb l1a$ depends only on the magnitude of $a$, we can, without loss
of generality, take $a\ge 0$ in this section. In the limit $z\to+\infty$,
formula (3.2.21) of Bateman \cite{Batema53} may be used to derive the following
asymptotic expression for $j\subb l1a(z)$
\begin{eqalignno}
j\subb l1a&=(2z)^{a-1}\frac{\Gamma(a)}{\Gamma(l+a+1)}
  F\biggl(\frac{l-a+1}2,\frac{-l-a}2;-a+1;-\frac1{z^2}\biggr)\nonumber\\
&{}+(2z)^{-a-1}\frac{\Gamma(-a)}{\Gamma(l-a+1)}
  F\biggl(\frac{l+a+1}2,\frac{-l+a}2; a+1; -\frac1{z^2}\biggr).
\end{eqalignno}
This cannot be used directly when $a$ is an integer or $l\pm a$ is a negative
integer; in those cases, a limit must be taken. For $a$ and $l$ both
non-negative integers, the leading order behavior is found to be
\begin{equation}
j\subb l1a\to
\begin{cases}
  \displaystyle\frac{(a-1)!}{(a+l)!}(2z)^{a-1},&for $a\ne0$,\\
  \displaystyle\frac{\sinh^{-1}z-\sum_{k=1}^l k^{-1}}{l!z},&for $a=0$,
\end{cases}
\end{equation}
and
\begin{equation}
y\subb l1a\to
\begin{cases}
  \displaystyle\frac{(-1)^{l+1}(a-1)!}{(a-l-1)!}(2z)^{a-1},&for $l<a$,\\
  \displaystyle\frac{2(a+l)!}{a!}(2z)^{-a-1},&for $l\ge a$.
\end{cases}
\end{equation}
For $0\le l<a$ and $a$ an integer, $j\subb l1a(z)$ and $y\subb l1a(z)$ have the
same behavior as $z\to+\infty$ (namely $z^{a-1}$). It is, therefore, convenient
to define some combination of $j\subb l1a$ and $y\subb l1a$ in which the
leading-order behavior cancels. Such a combination is
\begin{eqalignno}\label{q-def}
q\subb l1a(z)
  &=j\subb l1a(z)-(-1)^{l+1}\frac{(a-l-1)!}{(a+l)!}y\subb l1a(z)\nonumber\\
  &=-i\sqrt{\frac2{\pi z}}Q^{-l-1/2}_{a-1/2}(\gamma).
\end{eqalignno}
For $l<-a$, we have $q\subb l1a = j\subb l1a$; for $l\ge a$, $q\subb l1a$
diverges. The asymptotic behavior of $q\subb l1a$ for $l<a$ is
\begin{equation}
q\subb l1a\to\frac{(-1)^{l+1}2(a-l-1)!}{a!}(2z)^{-a-1}.
\end{equation}

An asymptotic series for $j\subb l1a(z)$ valid for $l\to+\infty$ follows from
formula (3.2.14) of Bateman \cite{Batema53}
$$
j\subb l1a(z)=\frac1{(2l+1)!!}\biggl(\frac2{\gamma+1}\biggr)^{l+1/2}z^l
  F\biggl(-a+\frac12,a+\frac12;l+\frac32;-\frac{z^2}{\gamma+1}\biggr).
$$
The leading term is
\begin{equation}\label{l-asympt}
j\subb l1a(z)\to\frac{\sqrt{\pi/2}}{\Gamma(l+3/2)}z^l(\gamma+1)^{-l-1/2},
 \qquad \mbox{for $l\to+\infty$}.
\end{equation}

\subsection{Recurrence relations}

Recurrence relations that give $j\subb l1a$ in terms of $j\subb {l\pm1}1a$ and
$j\subb l1{a\pm1}$ may be derived from the corresponding relations for Legendre
functions \cite{GraRyz65,Batema53}:
\begin{subequations}\label{rec-eqs}
\begin{eqalignno}
j\subb l1a(z)&=\frac{z}{(2l+1)\gamma}\bigl[j\subb{l-1}1a(z)\nonumber\\
&\qquad\qquad+(l+1-a)(l+1+a)j\subb{l+1}1a(z)\bigr]\label{rec-1}\\
&=-\frac1{2a\gamma}\bigl[(l-a+1)j\subb l1{a-1}(z)
-(l+a+1)j\subb l1{a+1}(z)\bigr]  \label{rec-4}\\
&=\frac1{(l-a)\gamma}\bigl[zj\subb{l-1}1a(z)-(l+1+a)j\subb l1{a+1}(z)\bigr].
  \label{rec-5}
\end{eqalignno}
\end{subequations}
These relations may be combined to express $j\subb l1a$ in terms of any pair of
its neighbors. The derivative of $j\subb l1a$ may be found from
\begin{equation}\label{rec-2}
\frac d{dz}j\subb l1a(z)=
  \frac1{\gamma}j\subb{l-1}1a(z)-\frac{l+1}{z}j\subb l1a(z).
\end{equation}
By combining Eqs.~(\ref{high-def}) with the relations~(\ref{rec-1}) and
(\ref{rec-2}), we can generalize these recurrence relations to multiple indices
\begin{eqalignno}
j\subb lk\s(z)&=j\subb{l-2}{k+1}{\s a}(z)
  -(2l-1)\frac{\gamma}z j\subb{l-1}{k+1}{\s a}(z)\nonumber\\
&\qquad\qquad+(l^2-a^2)j\subb l{k+1}{\s a}(z),\label{rec-eqs-a}\\
\frac d{dz}j\subb lk\s(z)&=
  \frac1{\gamma}j\subb{l-1}k\s(z)-\frac{l+1}{z}j\subb lk\s(z).\label{rec-2-a}
\end{eqalignno}
Other recurrences involving the higher order functions may be found by
differentiating Eq.~(\ref{rec-5}) with respect to $a$:
\begin{subequations}\label{rec-eqs-b}
\begin{eqalignno}
&2(a+1)(l+1+a)j\subb l2{a+1,a+1}(z)\nonumber\\
&\qquad{}=2a[zj\subb {l-1}2{a,a}(z)-(l-a)\gamma j\subb l2{a,a}(z)]\nonumber\\
&\qquad\qquad{}  +\gamma j\subb l1a(z)-j\subb l1{a+1}(z),\label{rec-6}\\
&4(a+1)^2(l+1+a)j\subb l3{a+1,a+1,a+1}(z)\nonumber\\
&\qquad{}=4a^2[zj\subb{l-1}3{a,a,a}(z)-(l-a)\gamma j\subb l3{a,a,a}(z)]
  +zj\subb {l-1}2{a,a}(z)\nonumber\\
&\qquad\qquad{}-(l-3a)\gamma j\subb l2{a,a}(z)
  -(l+3a+3)j\subb l2{a+1,a+1}(z).\label{rec-7}
\end{eqalignno}
\end{subequations}

Equations~(\ref{rec-eqs}) and (\ref{rec-2}) constitute a set of
second-order recurrence relations which $j\subb l1a$ solve. There is
another independent solution which we will write as $g\subb l1a$.
If we substitute
$$g\subb l1a(z)=c_{l,a} y\subb l1a(z)$$
into the recurrence relations, the resulting relations are equivalent to the
original set provided $c_{l,a}$ satisfies
\begin{eqalignno*}
\frac{c_{l,a}}{c_{l-1,a}} &= \frac1{(l+a)(l-a)},\\
\frac{c_{l,a}}{c_{l,a-1}} &= -\frac{l+1-a}{l+a}.
\end{eqalignno*}
Depending on the value of $l$, $c_{l,a}$ may be written in one of three
forms
\begin{equation}
c_{l,a}=\begin{cases}
\displaystyle c^{i}_{l,a}=\frac{(-1)^a}{(l-a)!(l+a)!},  &for $l\ge a$,\\
\displaystyle c^{ii}_{l,a}=\frac{(-1)^l(a-l-1)!}{(l+a)!},  &for $-a\le l<a$,\\
\displaystyle c^{iii}_{l,a}=(-1)^a(a-l-1)!(-a-l-1)!,  &for $l<-a$,
\end{cases}
\end{equation}
where we have taken $a\ge0$.
Thus $g^{ii}\subb l1a(z)=c^{ii}_{l,a}y\subb l1a(z)$ solves the recurrence
relations for $-a\le l<a$. If this solution is extended beyond this range, it
degenerates to 0 (for $l<-a$) or infinity (for $l\ge a$). The independence of
$j\subb l1a$ and $g\subb l1a$ may be verified by computing the Casoratians.
These may be found by substituting recurrence relations~(\ref{rec-eqs}) and
(\ref{rec-2}) into the Wronskian, Eq.~(\ref{wronskian}) to give
\begin{eqalignno*}
g\subb l1a(z) j\subb {l-1}1a(z) - j\subb l1a(z) g\subb {l-1}1a(z)&=
  -c_{l,a}\frac{1}{z^2},\\
g\subb l1a(z) j\subb l1{a-1}(z) - j\subb l1a(z) g\subb l1{a-1}(z)&=
  -\frac{c_{l,a}}{l+1-a}\frac{1}{z}.
\end{eqalignno*}

The function $q\subb l1a(z)$ defined in Eq.~(\ref{q-def}) may be
written as $j\subb l1a(z)-g^{ii}\subb l1a(z)$. Thus $q\subb l1a(z)$ solves
the recurrence relations for $l<a$, and this gives a solution independent
of $j\subb l1a$ for $-a\le l<a$.

\subsection{Generating function}

From the differential recurrence, Eq.~(\ref{rec-2-a}), we obtain
$$
f\subb {l+m}k\s(\gamma)=(-1)^m\frac{d^m}{d\gamma^m}f\subb lk\s(\gamma),
$$
where
$$
f\subb lk\s(\gamma)=\frac{y\subb lk\s(\sqrt{\gamma^2-1})}{(\gamma^2-1)^{l/2}}.
$$
Performing a Taylor-series expansion of $f\subb 0k\s(\gamma-\epsilon z)$
gives
$$
f\subb 0k\s(\gamma-\epsilon z)=
  \sum_{l=0}^\infty f\subb lk\s(\gamma)\frac{(\epsilon z)^l}{l!}.
$$
This leads to the following generating function
\begin{equation}\label{gen-fun}
y\subb 0k\s\bigl(\sqrt{z^2(1+\epsilon^2)-2\epsilon z\gamma}\bigr)
  =\sum_{l=0}^\infty y\subb lk\s(z)\frac{\epsilon^l}{l!}.
\end{equation}

\subsection{Special cases}

Simple closed expressions exist in certain special cases (in the following
equations, $\sigma=\sinh^{-1}z$):
\begin{subequations}
\begin{eqalignno}
j\subb a1a(z)&=\frac1{(2a-1)!!z^{a+1}}\int_0^z\frac{z'^{2a}}{\gamma'}\,dz',\\
j\subb {a-1}1a(z)&=\frac{z^{a-1}}{(2a-1)!!},\label{spec-c}\\
j\subb 01a(z)&=
  \frac{\sinh(a\sigma)}{az}=\frac{(\gamma+z)^a-(\gamma+z)^{-a}}{2az},
  \label{spec-a}\\
j\subb{-1}1a(z)&=
  \frac{\cosh(a\sigma)}z=\frac{(\gamma+z)^a+(\gamma+z)^{-a}}{2z},
  \label{spec-b}\\
j\subb{-a-1}1a(z)&=(-1)^a\frac{(2a-1)!!}{z^{a+1}},\label{spec-d}\\
j\subb 0k{0\ldots0}(z)&=\frac{\sigma^{2k-1}}{(2k-1)!z},\label{start}
\end{eqalignno}
\end{subequations}
where we have taken $a\ge0$.

The recursion relations, Eqs.~(\ref{rec-eqs}), (\ref{rec-eqs-a}), and
(\ref{rec-eqs-b}), together with
Eq.~(\ref{start}) allow $j\subb lk\s$ for $0\le k\le3$ for
integer $l$ and integer indices $\s$ to be expressed in terms of elementary
functions. The multiple-index homogeneous solutions that we need can be
expressed simply in terms of the single-index solutions:
\begin{subequations}\label{mult-index}
\begin{eqalignno}
j\subb l2{02}(z)&=\frac z2 j\subb{l+1}11,\\
j\subb l2{11}(z)&=\frac z2 j\subb{l+1}10,\\
j\subb l2{22}(z)&=\frac z2 j\subb{l+1}11+\frac{z^2}2 j\subb{l+2}10,\\
j\subb l3{022}(z)&=\frac {z^2}8 j\subb{l+2}10.
\end{eqalignno}
\end{subequations}
For definiteness, we catalog all the required functions $j\subb lk\s(z)$ and
$y\subb lk\s(z)$ for $l=0$ and 1:
\begin{subequations}\label{l=0}
\begin{eqaligntwo}
j\subb010 &= \frac{\sigma}{z},&y\subb010 &= -\frac1z,\\
j\subb011 &= 1,   &y\subb011 &= -\frac\gamma z,\\
j\subb012 &= \gamma,&y\subb012 &= -\frac{1+2z^2}{z},
\end{eqaligntwo}
\begin{eqaligntwo}
j\subb02{02} &= \frac{z\gamma-\sigma}{4z},&
y\subb02{02} &= -\frac z2,\\
j\subb02{11} &= \frac{\gamma\sigma-z}{2z},&
y\subb02{11} &= -\frac \sigma 2,\\
j\subb02{22} &= \frac{-z\gamma+(1+2z^2)\sigma}{8z},&
y\subb02{22} &= -\frac {\gamma\sigma}2,
\end{eqaligntwo}
\begin{eqaligntwo}
j\subb03{022} &= \frac{-3z\gamma+(3+2z^2)\sigma}{32z},&
y\subb03{022} &= \frac {-\gamma\sigma+z}8.
\end{eqaligntwo}
\end{subequations}
\begin{subequations}\label{l=1}
\begin{eqaligntwo}
j\subb110 &= \frac {\gamma\sigma-z}{z^2},&
  y\subb110 &= -\frac \gamma{z^2},\\
j\subb111 &= \frac{z\gamma-\sigma}{2z^2},&
  y\subb111 &= -\frac 1{z^2},\\
j\subb112 &= \frac z3,&
  y\subb112 &= -\frac{(1-2z^2)\gamma}{z^2},
\end{eqaligntwo}
\begin{eqaligntwo}
j\subb12{02} &= \frac{-3\gamma\sigma+(3z+z^3)}{12z^2},&
y\subb12{02} &= \frac \gamma 2, \label{l=1-ex}\\
j\subb12{11} &= \frac{-3z\gamma+(3+2z^2)\sigma}{8z^2},&
y\subb12{11} &= \frac 12,\\
j\subb12{22} &= \frac{-(3-6z^2)\gamma\sigma+(3z-5z^3)}{72z^2},&
y\subb12{22} &= \frac {\gamma+z\sigma}2,
\end{eqaligntwo}
\begin{eqaligntwo}
j\subb13{022} &= \frac{(15+6z^2)\gamma\sigma-(15z+11z^3)}{288z^2},&
y\subb13{022} &= \frac {z\sigma}8.
\end{eqaligntwo}
\end{subequations}
The function $y\subb 0k\s(z)$ is the kernel that occurs in
Eq.~(\ref{pot-int}).

\subsection{Numerical methods}

Our final concern is to determine a method by which $j\subb lk\s$,
$y\subb lk\s$, and their deriv\-atives may be calculated accurately and
quickly. A direct use of the analytic forms is ill-advised, particularly when
$z$ is small and $l$ is large. For example, consider the numerator of the
analytic form for $j\subb 12{02}$ given in Eq.~(\ref{l=1-ex}) in the limit
$z\to0$. This consists of three terms, the largest of which is proportional
to $z$; however, the sum is proportional to $z^5$.

Given $j\subb l1a$, Eqs.~(\ref{mult-index}) may be used to calculate
the required multiple-index solutions. Furthermore
Eq.~(\ref{rec-2-a}) may be used to give the derivatives of $j\subb
lk\s$. Equation~(\ref{A-y-def}) gives $y\subb lk\s$ in terms of
$j\subb lk\s$.
Thus the problem is reduced to calculating $j\subb l1a(z)$ of
all integer $-L-1\le l\le L$, integer $a\ge 0$, and real $z\ge0$. (For
$a<0$, we can use $j\subb l1{-a}(z) =j\subb l1a(z)$. For $z<0$,
we can use $j\subb l1a(-z) = (-1)^lj\subb l1a(z)$.)

We will also be able to avoid problems with numerical underflow and
overflow by computing $\jb la$ where
$$j\subb l1a(z) = \frac{z^l}{(2l+1)!!} \jb la(z).$$
In the nonrelativistic limit $z\to0$, we have, from
Eq.~(\ref{taylor-1}), $\jb la\to 1$.

Our main tool for calculating $j\subb l1a$
will be the recurrence relation~(\ref{rec-1}) which when written in terms
of $\jb la$ becomes
\begin{equation}\label{rec-n}
\jb {l-2}a = \gamma\jb {l-1}a - \frac{(l-a)(l+a)}{(2l-1)(2l+1)}z^2 \jb la.
\end{equation}
In order to apply it we
need to examine the stability of the recurrence for large $l$. In this
limit, the recurrence relation is approximately
$$
\jb {l-2}a \approx \gamma\jb {l-1}a - \sfrac14 z^2 \jb la,
$$
whose solution is
$$
\frac{\jb la}{\jb{l-1}a} \approx
\frac{2}{\gamma\pm 1}.
$$
Comparing this with the leading term in the asymptotic
series~(\ref{l-asympt}) shows that the solution we want corresponds to the
upper sign. This solution is dominant when the recurrence relation is
applied in the backwards direction. If we start with large $l$ with some
arbitrary mixture of the dominant and subdominant solutions, then on each
application of the backwards recurrence relation, the subdominant solution
decreases by $(\gamma-1)/(\gamma+1)=z^2/(\gamma+1)^2$ relative to the
desired solution.

If we desire to compute $\jb la$ to accuracy
$\delta$ for $l\le L$, we choose an $L'$ such that
$$L'>L+\frac{\log(\delta)}{2\log(z/(\gamma+1))}.$$
We set $\jb {L'}a'$ and $\jb {L'-1}a'$ so that their ratio is given by
$\jb {L'}a'/\jb{L'-1}a' = 2/(\gamma+1)$. We then use
Eq.~(\ref{rec-n}) as a backwards recurrence to give $\jb la'$ for
$0\le l\le L$. At this point $\jb la'$ differs from the desired
solution only by an overall multiplicative factor. This may be determined
from Eq.~(\ref{spec-c}) which gives $\jb {a-1}a  = 1$. Thus $\jb la =
\jb la'/\jb {a-1}a'$. Because of the degeneracy in Eq.~(\ref{rec-n})
the values of $\jb la$ for $0\le l<a$ are independent of the choice of $L'$
and starting values $\jb{L'}a'$ and $\jb{L'-1}a'$. The recurrence is
effectively restarted at $l=a-1$.

Various optimizations to this scheme are possible. For example, it is only
necessary to start the recursion at $l=L'$ for one value of $a$, e.g.,
$a=0$. For other values of $a$ we can start the recursion at $l=L$ by
rewriting Eq.~(\ref{rec-5}) as
\begin{equation}\label{rec-nn}
\jb la =\frac{(2l+1)\jb{l-1}{a-1}-(l+1-a)\gamma\jb l{a-1}}{l+a},
\end{equation}
and using this recurrence to give $\jb La$ and $\jb {L-1}a$ in terms of
$\jb L{a-1}$, $\jb {L-1}{a-1}$, and $\jb {L-2}{a-1}$.

For large values of $z$, $L'$ becomes large because the behavior of the
dominant and subdominant solutions is nearly the same. It is, therefore,
possible to use forward recursion using Eq.~(\ref{rec-n}) to obtain
$\jb la$ for $l>a$. Starting values are given by $\jb{a-1}a = 1$ and $\jb
aa$ which may be calculated using $\jb 00 = \sigma/z$ and recurrence
relation~(\ref{rec-nn}). As before, backwards recursion should be used for
$0\le l <a$.

For $l<-a$, we can compute $\jb la$ by backwards recursion using
Eq.~(\ref{rec-n}) together with the starting value $\jb {-a-1}a =1$.
For $-a\le l<0$, we could continue the backwards recursion using as
starting values $\jb 0a$ and $\jb 1a$ as found above. However, it is
sometimes useful to be able to compute $q\subb l1a$; but this cannot be
accurately computed using Eq.~(\ref{q-def}) when $z$ is large
because of the large cancellation that occurs in this limit. Instead, we
compute $q\subb l1a$ directly. To avoid problems with underflow and
overflow, we work with $\qb la$ which is defined by
$$q\subb l1a(z) = \frac{z^l}{(2l+1)!!} \qb la(z).$$
We have seen that $q\subb l1a$ satisfies the same recurrence relations as
$j\subb l1a$. This implies that $\qb la$ satisfies the recurrence
relation~(\ref{rec-n}). From Eqs.~(\ref{spec-a}) and (\ref{spec-b}),
together with the recurrence relation, we have
\begin{subequations}
\begin{eqalignno}
\qb{-1}a(z) &=\frac1{(\gamma+z)^a},\\
\qb{-2}a(z) &=\frac{\gamma+az}{(\gamma+z)^a}.
\end{eqalignno}
\end{subequations}
We can then utilize backwards recursion using
Eq.~(\ref{rec-n})
to give $\qb la$ for all $-a\le l<0$. Finally, we can compute $\jb la$ for
$-a\le l<0$ using Eq.~(\ref{q-def}) which gives
\begin{equation}\label{q-j}
\jb la(z)=\qb la(z) + \frac{(a-l-1)!}{(a+l)!}
\frac{(-1)^{l+1}z^{-2l-1}}{(-2l-3)!!(-2l-1)!!}
\jb {-l-1}a.
\end{equation}

\bibliographystyle{\bibliostyle}
\bibliography{cond}
\begin{thetables*}{I}
\tableitem{cond-tab} Conductivities for various values of the normalized
temperature $\Theta = T_e/\linebreak[0](511\,\rm keV)$ and the effective
ion charge state $Z$. The conductivities are normalized to
$(4\pi\epsilon_0^2)/(m_e^{1/2}q_e^2\log\Lambda^{e/e})(T_e^{3/2}/Z)$.
% results from [coll]spitz.for.82; July 19, 1988
% calculated with dd=20, n=2000, iter2=200, eps=1.e-9, lin=.t, correct=.t
% except iter2=0 for zi=0 and lin=.f for zi=inf.

\begin{center}
\def~{\phantom{0}}
\unitlength=3em\def\arraystretch{0}\tabcolsep=5.5pt
\begin{tabular}
{|@{\vrule height2.75ex depth.75ex width0em}c@{}|cccccc|}
\hline\hline
\raisebox{-.4\unitlength}
{\begin{picture}(1.25,1)(-1.25,0)
\put(0,0){\line(-1,1){1}}
\put(-.75,.3){\makebox(0,0){$\Theta$}}
\put(-.25,.75){\makebox(0,0){$Z$}}
\end{picture}}
       &  0~     &  1       &  2       &  5       & 10       & $\infty$ \\
\hline
0      & 3.75994 & ~7.42898 & ~8.75460 & 10.39122 & 11.33006 & 12.76615 \\
0.01   & 3.75490 & ~7.27359 & ~8.53281 & 10.07781 & 10.95869 & 12.29716 \\
0.02   & 3.74920 & ~7.12772 & ~8.32655 & ~9.78962 & 10.61952 & 11.87371 \\
0.05   & 3.72852 & ~6.73805 & ~7.78445 & ~9.04621 & ~9.75405 & 10.81201 \\
0.1    & 3.68420 & ~6.20946 & ~7.06892 & ~8.09361 & ~8.66306 & ~9.50746 \\
0.2    & 3.57129 & ~5.43667 & ~6.06243 & ~6.80431 & ~7.21564 & ~7.82693 \\
0.5    & 3.18206 & ~4.13733 & ~4.47244 & ~4.88050 & ~5.11377 & ~5.47602 \\
1      & 2.65006 & ~3.13472 & ~3.32611 & ~3.57303 & ~3.72206 & ~3.96944 \\
2      & 2.03127 & ~2.27862 & ~2.39205 & ~2.54842 & ~2.64827 & ~2.82473 \\
5      & 1.33009 & ~1.45375 & ~1.51805 & ~1.61157 & ~1.67382 & ~1.78870 \\
10     & 0.94648 & ~1.02875 & ~1.07308 & ~1.13856 & ~1.18263 & ~1.26490 \\
20     & 0.67042 & ~0.72743 & ~0.75853 & ~0.80472 & ~0.83593 & ~0.89443 \\
50     & 0.42422 & ~0.46003 & ~0.47965 & ~0.50885 & ~0.52861 & ~0.56569 \\
100    & 0.29999 & ~0.32528 & ~0.33915 & ~0.35979 & ~0.37377 & ~0.40000 \\
\hline\hline
\end{tabular}
\end{center}

\end{thetables*}

\begin{thefigures*}{9}
\figitem{fig-1} The normalized conductivity as a function of the
normalized temperature $\Theta = T_e/(511\,\rm keV)$ for various values of the
effective ion charge state $Z$.
\begin{figure}
\begin{center}
\mbox{ }\\
\epsfxsize=4.5in
\epsffile{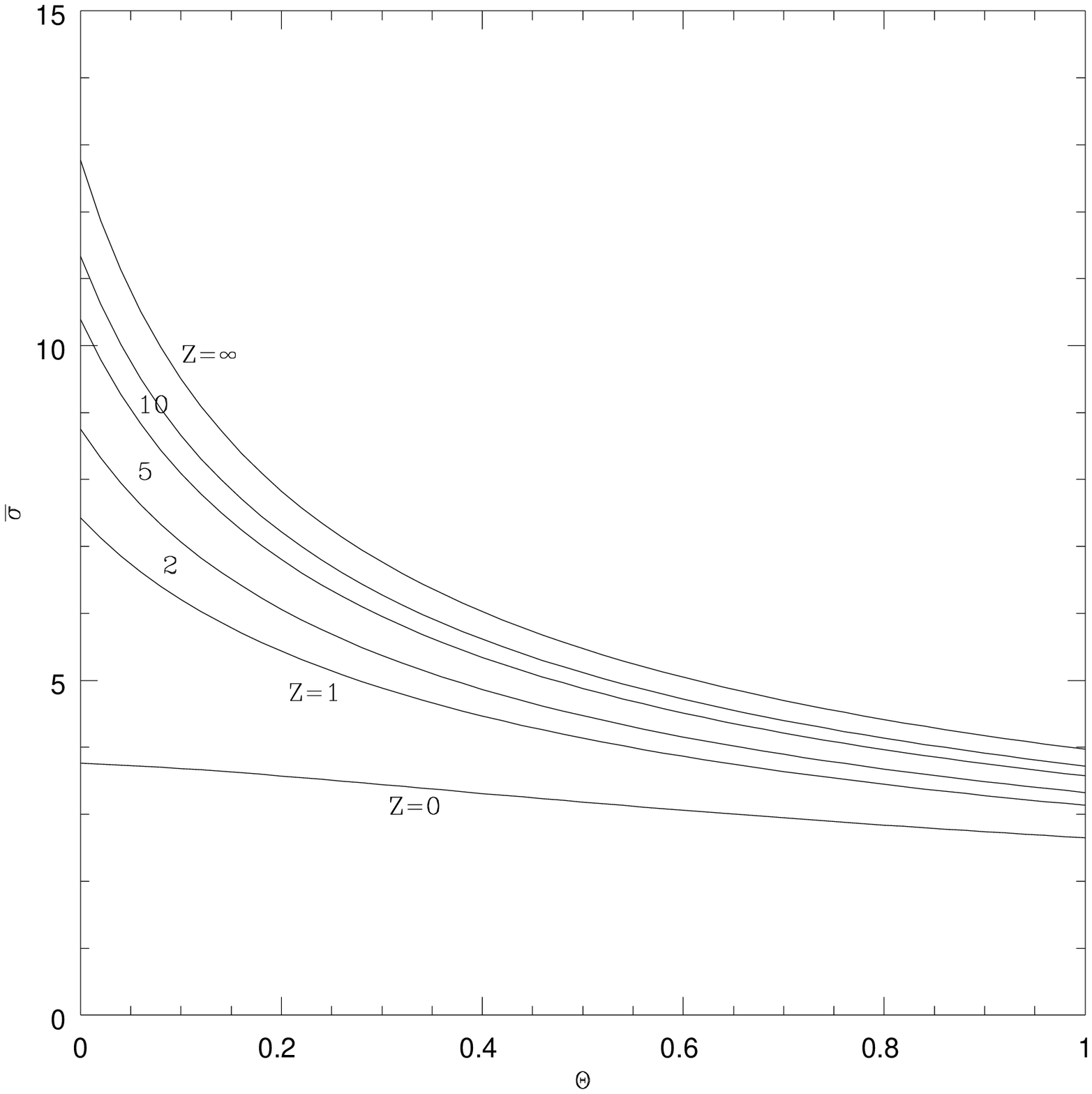}
\end{center}
\end{figure}
\end{thefigures*}
\end{document}